\documentclass[a4paper,usenatbib]{mnras}

\usepackage{graphicx} % package to manage images
\usepackage[usenames,dvipsnames]{color}
\usepackage{latexsym}
\usepackage{amssymb}
\usepackage{amsmath}
\usepackage{soul}
\usepackage{ae,aecompl}

\usepackage{hyperref}
\hypersetup{
    colorlinks=true,
    linkcolor=blue,
    filecolor=blue,      
    urlcolor=red,
    citecolor=blue,
}

\title[Ex-situ discs in Auriga]{Lessons from the Auriga discs: The hunt for the Milky Way's ex-situ disc is
not yet over}
\author[F. A. Gomez et al.]{Facundo A. G\'omez$^{1}$\thanks{E-mail: fgomez@mpa-garching.mpg.de}, 
Robert J. J. Grand$^{2,3}$,
Antonela Monachesi$^{1}$, \newauthor
Simon D. M. White$^{1}$, 
Sebastian Bustamante$^{2}$,
Federico Marinacci$^{4}$,\newauthor 
R{\"u}diger Pakmor$^{2}$, 
Christine M. Simpson$^{2}$,
Volker Springel$^{2,3}$, 
and Carlos S. Frenk$^{5}$\\
$^{1}$Max-Planck-Institut f\"ur Astrophysik, Karl-Schwarzschild-Str. 1, D-85748, Garching, Germany\\
$^{2}$Heidelberger Institut f\"ur Theoretische Studien, Schloss-Wolfsbrunnenweg 35, 69118 Heidelberg, Germany\\
$^{3}$Zentrum f\"ur Astronomie der Universitat Heidelberg, Astronomisches Recheninstitut, Monchhofstr. 12-14, 69120 Heidelberg, Germany\\
$^{4}$Department of Physics, Kavli Institute for Astrophysics and Space Research, MIT, Cambridge, MA 02139, USA\\
$^{5}$Institute for Computational Cosmology, Department of Physics, Durham University, South Road, Durham, DH1 3LE, UK\\
}

\begin{document}

\date{}

\pagerange{\pageref{firstpage}--\pageref{lastpage}} \pubyear{2017}

\maketitle

\label{firstpage}

\begin{abstract}
We characterize the contribution from accreted material to the galactic discs of the Auriga Project, 
a set of high resolution magnetohydrodynamic cosmological simulations of late-type galaxies performed with the moving-mesh code 
{\sc AREPO}.  
Our goal is to explore whether a significant accreted (or ex-situ)  stellar component in the Milky Way disc could be hidden within the 
near-circular orbit population, which is strongly dominated by stars born in-situ. One third of our models shows a significant ex-situ disc
but this fraction would be larger if constraints on orbital circularity were relaxed.
Most of the ex-situ material ($\gtrsim 50\%$) comes from single massive satellites 
($> 6 \times 10^{10}~M_{\odot}$). These satellites are accreted with a wide range of infall times and inclination angles (up to $85^{\circ}$).  
Ex-situ discs are thicker, older and more metal-poor than their in-situ counterparts. They show a flat median age profile, 
which differs from the negative gradient observed in the in-situ component. As a result, the likelihood of
identifying an ex-situ disc in samples of old stars on near-circular orbits increases towards the outskirts of the disc. 
We show three examples that, in addition to ex-situ discs, have a strongly rotating dark matter component.  
Interestingly, two of these ex-situ stellar discs show an orbital circularity distribution that is consistent with that of the in-situ disc. 
Thus, they would not be detected in typical kinematic studies. 

\end{abstract}


\section{Introduction}
\label{sec:intro}

According to the currently favored cosmological model, $\Lambda$ cold
dark matter ($\Lambda$CDM), galaxies like our own merge and interact
with companions of widely different masses throughout their history
\citep{springel2008b}. The quantification and characterization of the
merger activity that a galaxy has undergone can thus be used to
constrain galaxy formation models.

Mergers and interactions with very low mass satellites (masses $\lesssim 10^{7} 
M_{\odot}$) are difficult to detect since such satellites are expected to possess, if any, only a small number of stars
\citep[see][and references therein]{2016MNRAS.456...85S}. The detection of truly ``dark'' satellites would be extremely 
rewarding as it could put stringent constraints on the nature of DM
\citep[e.g.][]{springel2008b,2012MNRAS.423.3740V,2013MNRAS.428..882M, 2016MNRAS.460..363L,2017MNRAS.464.4520B}. Nonetheless, 
an undisputed detection of this type of substructure is yet to be made. 

The identification of mergers associated with intermediate mass satellites (i.e. masses $< 10\%$ of the host mass)
is significantly less challenging. As such satellites interact with the host 
gravitational potential, they are tidally disrupted leaving behind debris in the form of stellar streams. Depending on the
time since disruption this debris can be detected either in real space, in the form of extended cold streams (recent disruption) 
\citep[e.g.][]{ibata94,2001ApJ...547L.133I,belu06,2007ApJ...658..337B}, 
or as clumps in the space of 
quasi-conserved integrals of motions (well after disruption) \citep[][]{hwzz99,2006MNRAS.365.1309H,klement09,morri09,2014ApJ...791..135H}. In the Milky Way several streams have been identified and, in some cases, 
their progenitors have been characterized. However, a robust quantification of our Galaxy's merger activity is still lacking. 
The main reason for this has been the lack of sufficiently large and accurate full phase-space catalogs that could unveil
debris from early accretion events, which would be generally be deposited in the inner Galactic regions. Thanks to the astrometric satellite {\it Gaia}
\citep{2016A&A...595A...2G}, in 
combination with previous and upcoming spectroscopic surveys \citep[e.g.][]{rave,2012Msngr.147...25G,lamost,2014SPIE.9147E..0MD,2016ASPC..507...97D}, this will soon
be possible \citep{hz00,gh10b}. Indeed, 
the first {\it Gaia} data release has started to uncover previously unknown substructure in the Galactic stellar halo \citep{2017A&A...598A..58H}. 

Isolating debris from more massive merger (i.e. masses $\geq 10\%$ of the host mass) is, however, more challenging. 
As discussed by \citet[][hereafter R14]{2014MNRAS.444..515R}, the 
reason for this is two-fold. First, dynamical friction is most efficient for these massive objects. They are quickly dragged to small radii
where the mixing time scales are short. Second, debris from these satellites is kinematically hotter than that from smaller 
mass objects and thus mixes faster. 
Relatively massive mergers can  be detected indirectly, by searching for perturbations in the local velocity field of the Galactic
disc, both in-plane \citep{2009MNRAS.396L..56M, 2012MNRAS.419.2163G} and perpendicular to the Galactic plane 
\citep{2013MNRAS.429..159G, 2014MNRAS.440.1971W}. Substructure in the local disc velocity field has already been identified 
\citep{2009MNRAS.396L..56M, 2012MNRAS.419.2163G, 2012ApJ...750L..41W}. However, perturbations from the Galactic bar or spiral arms
and debris from significantly less massive satellites may explain much of this substructure 
\citep[e.g.][]{2006MNRAS.365.1309H, 2006A&A...449..533A, 2008ApJ...685..261K, 2009ApJ...700L..78A, 2016MNRAS.461.3835M}. 

The addition of extra dimensions to the analysis, based on chemical abundances patterns and stellar ages, is thus crucial to 
isolate debris coming from individual progenitors \citep{2014ApJ...791..135H,2014ApJ...781L..20M,2014MNRAS.444..515R,2015MNRAS.450.2874R}. Indeed,
\citet[][hereafter R15]{2015MNRAS.450.2874R} analyzed a sample of $\sim 5000$ stars from the {\it Gaia}-ESO survey with full phase-space, [Mg/Fe]
and [Fe/H] measurements to search for signatures of the most massive merger events our Galaxy has undergone. Their efforts were focused on the
identification of an accreted or ex-situ disc component. Such an ex-situ disc is expected to arise during massive mergers at low
inclination angles with respect to the plane of the main disc 
\citep{2003ApJ...597...21A,2008MNRAS.389.1041R,2009MNRAS.397...44R,2009ApJ...703.2275P,2014ApJ...784..161P,2015ApJ...799..184P,2016MNRAS.461L..56S}.
A DM disc may also form at the same time, which could have important consequences for direct DM  detection experiments 
\citep[see][and references therein]{2016MNRAS.461L..56S}. 
To select ex-situ disc star candidates, R15 used a chemodynamical template first introduced by 
R14. As they discussed, for [Fe/H] $> -1.3$, stellar populations of surviving 
dwarf galaxies generally have [Mg/Fe]$ < 0.3$, which is lower than typical MW stars at the same [Fe/H]. To further 
isolate ex-situ disc candidates, R14 focused on stars co-rotating with the disc on orbits with significant eccentricity. 
Debris from massive mergers is expected to lie preferentially on 
such orbits. Both R14 and R15 found no evidence of a significant prograde ex-situ disc component of this type.
They concluded that the MW has no significant ex-situ stellar disc, and thus possesses no significant DM disc 
formed by a merger.   

In this paper we study the formation of ex-situ stellar discs in the Auriga simulations, a set of 
high-resolution magneto-hydrodynamic simulations of disc galaxy formation from $\Lambda$CDM initial conditions \citep{2017MNRAS.tmp...82G}.
\emph{Our goal is to explore whether a significant ex-situ stellar disc component could be hidden within the near-circular orbit 
population, which is strongly dominated by in-situ stars}. In Section~\ref{sec:sims} we introduce the Auriga suite. 
We define our ex-situ discs in Section~\ref{sec:exsitu_def} and quantify the number of models with a significant 
such component in Section~\ref{sec:quanty}. In Section~\ref{sec:formation} we show how our ex-situ discs are 
formed, and we characterize their main 
stellar populations properties in Section~\ref{sec:charac}. In Section~\ref{sec:identy} we discuss how these discs might be detected with 
upcoming observational campaigns. We discuss the implications of our findings for a possible ex-situ stellar and DM discs
in our Galaxy in Section~\ref{sec:discussion}. We conclude in Section~\ref{sec:conclusions}.

\begin{table*}
\centering
\caption{Table of simulation parameters. The columns are 1) Model
  name; 2) Virial mass; 3) Virial radius; 4) Stellar mass; 5) Disc
  stellar mass; 6) Disc radial scale length; 7) Bulge stellar mass; 8)
  Bulge effective radius; 9) Sersic index of the bulge, 10) Disc to
  total mass ratio and 11) Optical radius. See GR17 for definitions.}
\label{t1}
\begin{tabular}{c c c c c c c c c c c}
\hline

Run & $M_{\rm vir}$ & $R_{\rm vir}$ & $M_{*}$ & $M_{\rm d}$ & $R_{\rm d}$ & $M_{\rm b}$ & $R_{\rm eff}$ & $n$ & $D/T$ & $R_{\rm opt}$ \\
 & $[\rm 10^{10} M_{\odot}]$ & $[{\rm kpc}]$ & $[\rm 10^{10} M_{\odot}]$ & $[\rm 10^{10} M_{\odot}]$ & ${\rm [kpc]}$ & $[\rm 10^{10} M_{\odot}]$ & ${\rm [kpc]}$ &  &  & $[{\rm kpc}]$ \\
\hline

     Au-2  &  191.466  &   261.757   &    7.045   &    6.377   &   11.644    &   2.341   &    2.056  &     1.529 &       0.73	&	37.0\\
     Au-3  &  145.777  &   239.019   &    7.745   &    6.288   &    7.258    &   2.039   &    1.488  &     0.987 &       0.76	&	31.0\\
     Au-4  &  140.885  &   236.310   &    7.095   &    3.662   &    3.929    &   2.005   &    1.740  &     1.352 &       0.65	&	24.5\\
     Au-5  &  118.553  &   223.091   &    6.722   &    4.509   &    3.583    &   1.806   &    0.839  &     0.874 &       0.71	&	21.0\\
     Au-6  &  104.385  &   213.825   &    4.752   &    3.315   &    5.949    &   1.649   &    2.851  &     2.000 &       0.67	&	26.0\\
     Au-7  &  112.043  &   218.935   &    4.875   &    2.458   &    5.140    &   2.045   &    1.731  &     1.740 &       0.55	&	25.0\\
     Au-8  &  108.062  &   216.314   &    2.990   &    2.457   &    6.572    &   0.652   &    2.147  &     1.328 &       0.79	&	25.0\\
     Au-9  &  104.971  &   214.224   &    6.103   &    3.597   &    3.367    &   2.169   &    0.999  &     0.948 &       0.62	&	19.0\\
     Au-10 &  104.710  &   214.061   &    5.939   &    2.378   &    2.596    &   3.152   &    1.080  &     1.181 &       0.43	&	16.0\\
     %Au-11 &  164.935  &   249.053   &    6.909   &    1.133   &    3.055    &   2.750   &    1.046  &     1.354 &       0.29	&	16.0\\
     Au-12 &  109.275  &   217.117   &    6.010   &    4.315   &    3.290    &   1.134   &    0.892  &     0.759 &       0.79	&	19.0\\
     Au-13 &  118.904  &   223.325   &    6.194   &    1.675   &    3.382    &   3.798   &    1.403  &     1.549 &       0.31	&	15.5\\
     Au-14 &  165.721  &   249.442   &   10.393   &    6.359   &    4.186    &   3.184   &    1.138  &     1.586 &       0.67	&	26.0\\
     Au-15 &  122.247  &   225.400   &    3.930   &    2.772   &    5.320    &   1.047   &    2.216  &     2.000 &       0.73	&	23.0\\
     Au-16 &  150.332  &   241.480   &    5.410   &    5.059   &    9.030    &   1.175   &    1.825  &     1.391 &       0.81	&	36.0\\
     Au-17 &  102.835  &   212.769   &    7.608   &    2.563   &    4.191    &   4.641   &    1.208  &     0.831 &       0.36	&	16.0\\
     Au-18 &  122.074  &   225.288   &    8.037   &    5.205   &    3.719    &   2.461   &    1.225  &     0.950 &       0.68	&	21.0\\
     Au-19 &  120.897  &   224.568   &    5.320   &    3.532   &    4.805    &   1.428   &    1.416  &     2.000 &       0.71	&	24.0\\
     Au-20 &  124.922  &   227.028   &    4.740   &    2.248   &    8.019    &   2.353   &    2.174  &     1.886 &       0.49	&	30.0\\
     Au-21 &  145.090  &   238.645   &    7.717   &    6.005   &    4.607    &   1.240   &    1.188  &     1.064 &       0.83	&	24.0\\
     Au-22 &   92.621  &   205.476   &    6.020   &    2.851   &    2.249    &   2.709   &    1.014  &     0.934 &       0.51	&	13.5\\
     Au-23 &  157.539  &   245.274   &    9.023   &    5.547   &    4.985    &   3.192   &    1.708  &     1.438 &       0.63	&	25.0\\
     Au-24 &  149.178  &   240.856   &    6.554   &    3.756   &    5.570    &   2.186   &    0.946  &     0.969 &       0.63	&	30.0\\
     Au-25 &  122.109  &   225.305   &    3.142   &    2.475   &    6.695    &   0.934   &    2.951  &     1.879 &       0.73	&	21.0\\
     Au-26 &  156.384  &   244.685   &   10.967   &    4.697   &    3.141    &   5.456   &    1.116  &     1.017 &       0.46	&	18.0\\
     Au-27 &  174.545  &   253.806   &    9.606   &    7.229   &    4.287    &   1.742   &    0.947  &     1.065 &       0.81	&	26.0\\
     Au-28 &  160.538  &   246.833   &   10.448   &    6.761   &    2.159    &   2.359   &    0.948  &     1.109 &       0.74	&	17.5\\

\hline
\end{tabular}
\end{table*}

\section{The Auriga Simulations}
\label{sec:sims}

In this paper we analyze a subsample of the cosmological magnetohydrodynamic simulations of the Auriga suite \citep[][hereafter GR17]{2017MNRAS.tmp...82G}.  
 In what follows we summarize the main
characteristics of these simulations. For a more detailed description we refer the reader to GR17.

The Auriga suite is composed of 30 high-resolution cosmological zoom-in simulations of the formation of  
late-type galaxies within  Milky Way-sized haloes. The  haloes were selected from a lower resolution dark matter only 
simulation from the Eagle Project \citep{2015MNRAS.446..521S}, a periodic box  of side 100 Mpc. Each halo was chosen to  have, 
at $z=0$, a virial mass in the range of $10^{12}$ -- $2 \times 10^{12}$ M$_{\odot}$ 
and to be more distant than nine times the virial radius from any other halo of mass more than $3\%$ of 
its own mass. Each halo was
run at multiple resolution levels. The
typical dark matter particle and gas cell mass resolutions for the simulations used in this work 
(Auriga level 4) are  $\sim 3
\times 10^{5}$ $\rm M_{\odot}$ and $\sim 5 \times  10^{4}$ $\rm M_{\odot}$, 
respectively.   The  gravitational
softening length used for stars and DM grows with scale  factor up to a  maximum of 369
pc,  after which it  is kept  constant in physical units.  The  softening length  of gas
cells  scales with the  mean radius  of the  cell, but is never allowed to drop below the 
stellar softening length. A  resolution  study  across  three resolution levels (GR17) shows that many
galaxy   properties,  such  as   surface  density   profiles,  orbital
circularity  distributions, star formation histories  and  disc  vertical  structures  are already well
converged at the resolution level used in this work.

The simulations  were carried  out using the  $N$-body +  moving-mesh,
magnetohydrodynamics  code {\sc AREPO} \citep{2010MNRAS.401..791S,2016MNRAS.455.1134P}.
 A   $\Lambda$CDM    cosmology,   with
parameters  $\Omega_{\rm  m} =  \Omega_{\rm  dm}  +  \Omega_{\rm b}  =
0.307$,  $\Omega_{\rm  b}  =  0.048$, $\Omega_{\Lambda}=  0.693$,  and
Hubble  constant $H_{0}  = 100~h$  km s$^{-1}$  Mpc$^{-1}$ =  67.77 km
s$^{-1}$ Mpc$^{-1}$, was adopted.  

The baryonic physics model used in these simulations is a slightly
updated  version of that in \citet{2014MNRAS.437.1750M}. It  follows a number of 
processes that play a  key role in the
formation  of late-type  galaxies, such  as gas  cooling/heating, star
formation, mass  return and  metal enrichment from  stellar evolution,
the  growth of  supermassive black  holes, and  feedback both from stellar
sources and  from black hole  accretion. In addition, magnetic fields
were implemented as described in \citet{2013MNRAS.432..176P}. The effect 
of these magnetic fields on the global evolution of the Auriga galaxies 
is discussed in detail in \citet{2017arXiv170107028P}. The parameters that regulate
the efficiency of each physical process were chosen by comparing the results
obtained in simulations of cosmologically representative regions to a
wide range of observations of the galaxy population
\citep[][]{2013MNRAS.436.3031V,2014MNRAS.437.1750M}. 

From now on, we will
refer to these simulations as Au-$i$, with $i$ enumerating the
different initial conditions. We will focus on the subset of 26 models that, 
at the present-day, show a well defined stellar disc. The main properties of each simulated
galaxy are listed in Table~\ref{t1}. The disc/bulge decomposition is
made by simultaneously fitting exponential and Sersic profiles to
the stellar surface face-on density profiles. A detailed description of how
these parameters were obtained is given in GR17.

\section{Ex-situ disc definition}
\label{sec:exsitu_def}

The goal of this work is to characterize the contribution from ex-situ formed 
(accreted)\footnote{In this work the terms ex-situ and accreted are used interchangeably.} 
material to the Auriga stellar
discs. In what follows we will designate as ex-situ all star particles formed within the potential well
of a self-bound satellite galaxy prior to its disruption. At the present-day, such star particles can either belong to the main host, after 
being tidally stripped from their progenitor, or they can still be bound to this progenitor. 
Conversely, following \citet{2015MNRAS.454.3185C}, all 
particles formed within the potential well of the main host halo will be referred to as in-situ star particles. Note that,  contrary to \citet{2015ApJ...799..184P}, this 
definition includes star particles formed within the host virial radius out of gas recently stripped from satellites. 
These stellar particles are not  found in the galactic disc and thus do not affect our results. 

\begin{figure}
    \centering
    \includegraphics[width=80mm,clip]{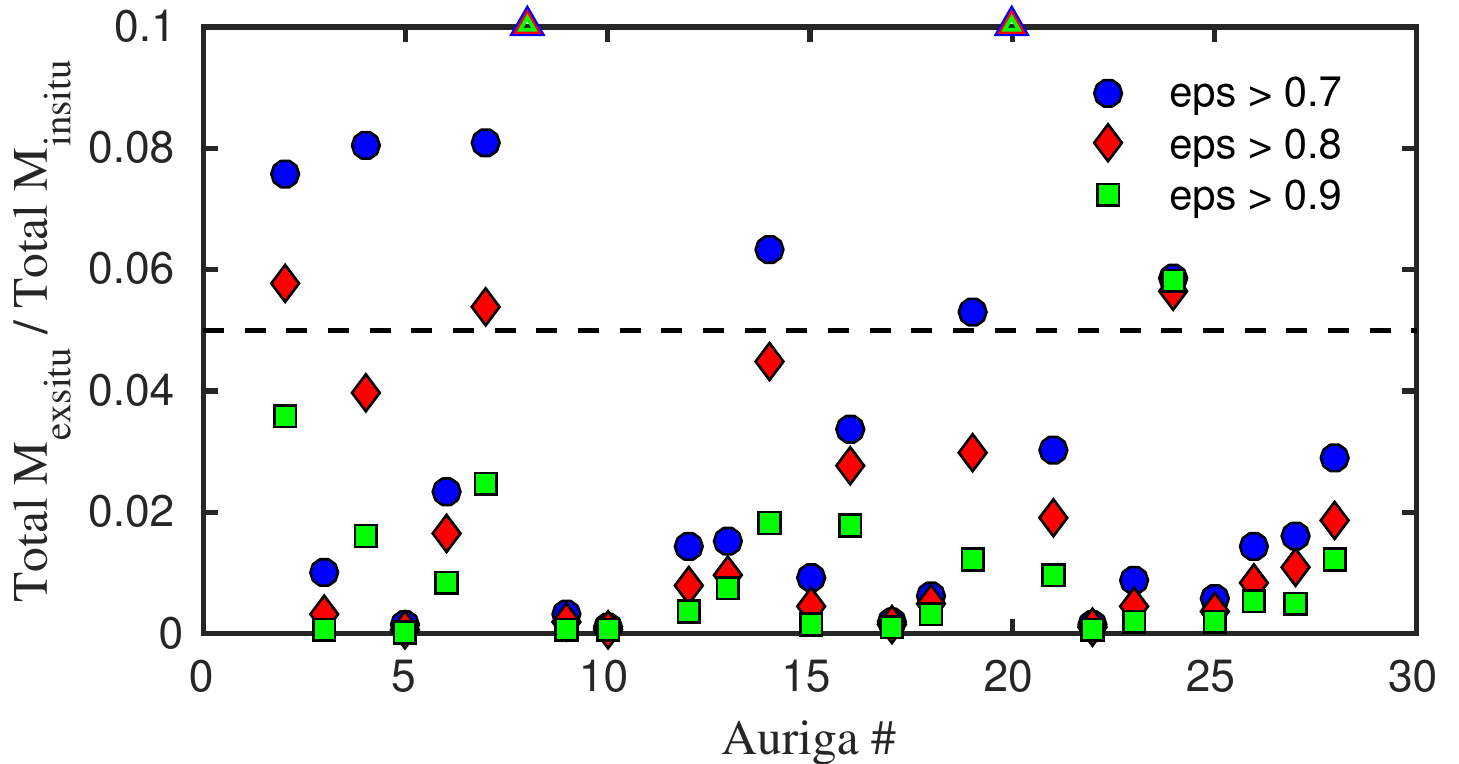}
    \caption{Ratio of the total ex-situ to in-situ disc mass, $\eta$, for the different Auriga galaxies. Only stellar particles with $R < R_{\rm opt}$ and
    $|Z| < 10$ kpc are considered. Blue, red and green symbols indicate the results obtained when a cut in the circularity parameter at 0.7, 0.8 and 0.9
    is imposed, respectively. The horizontal dashed lines indicate a $5\%$ mass ratio.}
    \label{fig:mass}
\end{figure}

To define star particles that are in the galactic disc we perform a kinematic decomposition based on the circularity parameter, $\epsilon$.
Following \citet{2003ApJ...597...21A}, this parameter is defined as

\begin{equation}
    \epsilon = \frac{L_{\rm z}}{L_{\rm z}^{\rm max}(E)},
\end{equation}
where $L_{\rm z}$ is the $Z$-component of angular momentum of a given star particle and $L_{\rm z}^{\rm max}(E)$ 
is the maximum angular momentum allowed for its orbital energy, $E$. Before computing the star particle's angular 
momentum, each galaxy is re-oriented as described in GR17, i.e., the $Z$-axis direction is defined through the 
semi-minor axis of the moment of inertia tensor of the star particles within $0.1 R_{200}$. 

All star particles that satisfy {\it i)} $\epsilon \geq 0.7$, {\it ii)} $\lvert Z \rvert < 10$ kpc and {\it iii)} $R < R_{\rm opt}$ 
are considered as disc star particles, independently of birth location. Here $R$ is the galactocentric cylindrical distance
and $R_{\rm opt}$ is the galactic  optical radius, defined as the $25~{\rm mag}~{\rm arcsec}^{-2}$ B-band isophotal radius
\citep{1998gaas.book.....B}.
To minimize contamination from stellar halo populations, for significantly lopsided galaxies $R_{\rm opt}$ is defined as 
the minimum cylindrical radius where  $\mu_{\rm B} = 25~{\rm mag}~{\rm arcsec}^{-2}$.

\section{Ex-situ disc quantification}
\label{sec:quanty}

In this Section we quantify the number of Auriga galaxies that possesses a significant ex-situ disc. To identify such discs  we show in Figure~\ref{fig:mass} the ratio of  
total ex-situ to  in-situ disc mass, $\eta = M_{\rm exsitu}^{\rm tot} / M_{\rm insitu}^{\rm tot}$, for all 
our galactic discs.
Interestingly, $31\%$ (8) of our disc sample show a significant ex-situ disc, defined as $\eta \gtrsim 0.05$ 
(dashed line). The two largest ex-situ discs, Au-8 and Au-20 have $\eta$ of $\sim 0.15$ and $\sim 0.3$, respectively.
In general, the value of $\eta$ rapidly decreases as we increase the circularity 
threshold from $\epsilon = 0.7$ to 0.9. This is not surprising since the orbital eccentricity of ex-situ disc 
stars is expected to be, on average, larger than that of their in-situ counterparts
\citep[see e.g.][]{2009MNRAS.400L..61S}.

\begin{figure*}
    \includegraphics[width=180mm,clip]{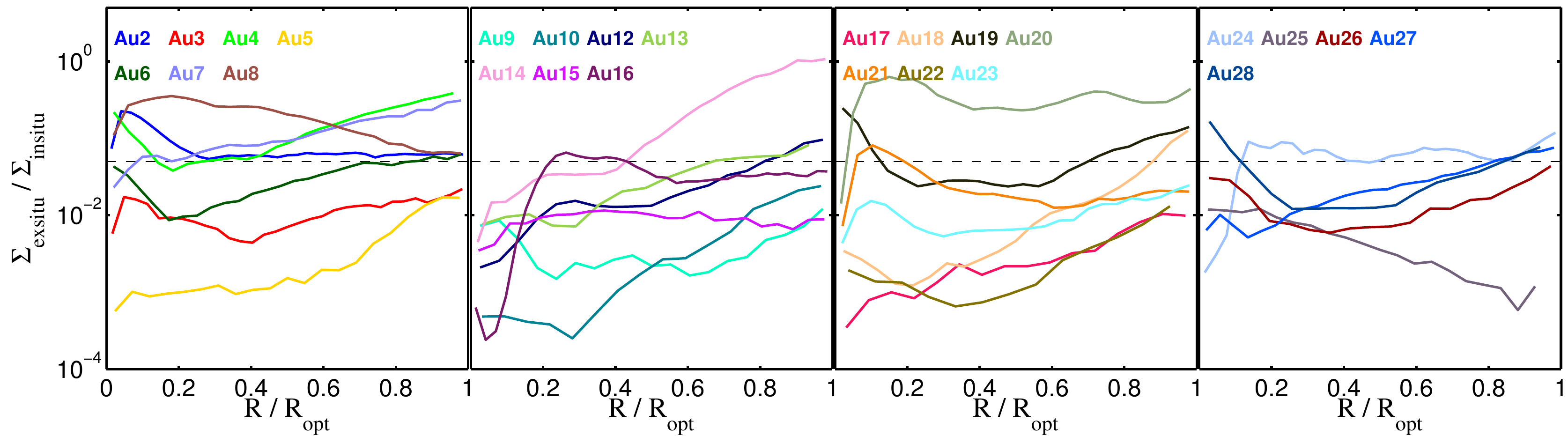}
    \caption{Surface density ratio, $\mu = \Sigma_{\rm exsitu}/\Sigma_{\rm insitu}$, as a function of galactocentric distance for the different Auriga 
    galaxies. For comparison, distances are normalized by the corresponding $R_{\rm opt}$. The horizontal dashed lines indicate a $5\%$ density ratio.}
    \label{fig:dens}
\end{figure*}

Figure~\ref{fig:dens} shows the surface density ratio, 
$\mu = \Sigma_{\rm exsitu} / \Sigma_{\rm insitu}$, as a function of galactocentric distance. 
To allow a direct comparison, distances are normalized by the corresponding $R_{\rm opt}$. Note that, in general, 
$\mu$ has a rising profile. This indicates that the relative contribution of ex-situ material to the disc rises 
as we move towards the outer disc regions. In the most extreme case, Au-14,  $\mu$ varies by approximately two orders of magnitudes within $R_{\rm opt}$.

Two-thirds of our disc sample shows either a very small or a negligible fraction of ex-situ  
material. Thus, in what follows, we will focus on the subsample of galaxies with significant ex-situ discs ($\eta > 0.05$). 

\begin{figure*}
\begin{flushleft}
    \includegraphics[width=170mm,clip]{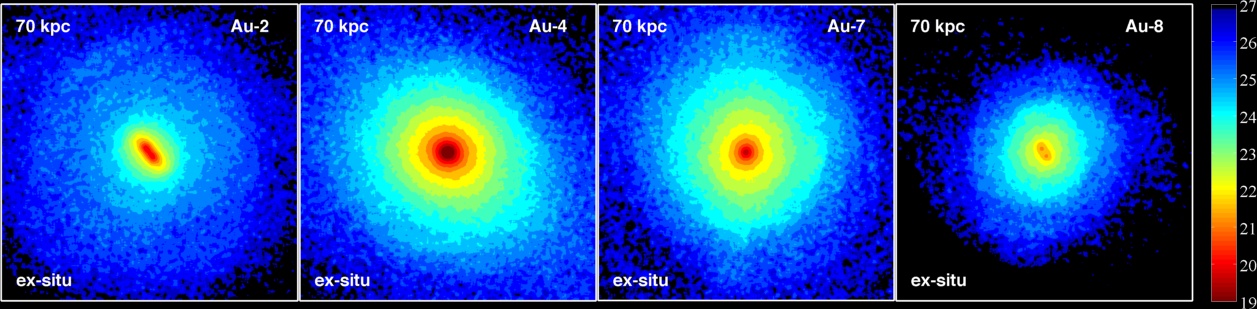}\\
    \includegraphics[width=170mm,clip]{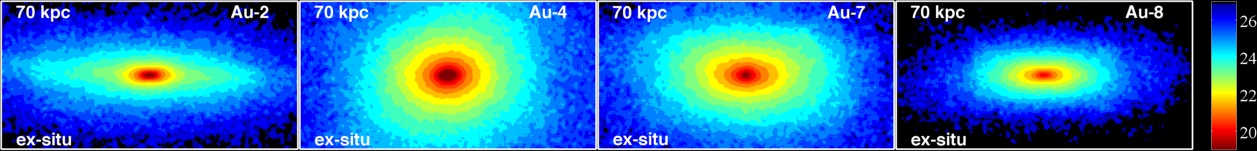}\\
    \includegraphics[width=170mm,clip]{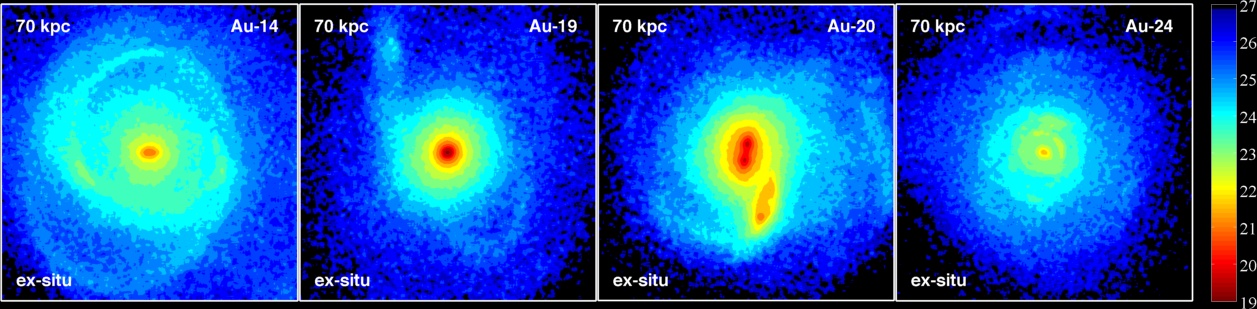}\\
    \includegraphics[width=170mm,clip]{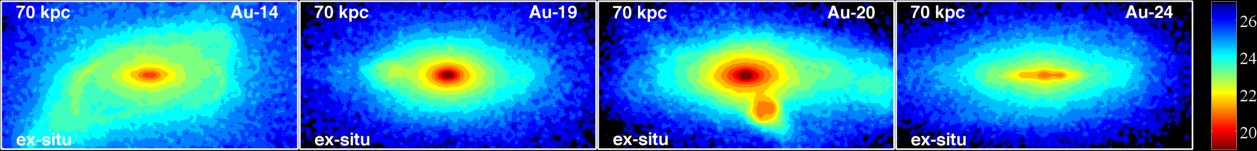}
    \end{flushleft}
    \caption{Present-day face-on and edge-on images of the B-band
      surface brightness of  galaxies that show a significant ex-situ
      disc. In this figure all ex-situ star particles are considered,
      independent of their circularity parameter. Only particles that
      belong to the main host are considered. The side length of each
      panel is 70 kpc. The colour bar indicates the scale for
      $\mu_{\rm B}$ in units of mag arcsec$^{-2}$.}
    \label{fig:sb_all}
\end{figure*}

Figure~\ref{fig:sb_all} shows the B-band surface brightness maps of these galaxies, obtained considering only 
ex-situ star particles.
In this figure we include all ex-situ star particles that at the present-day belong to the main host, 
independently of their circularity parameter. Thus, 
 stellar populations that belong to the galactic spheroid, or stellar halo, are also included. 
In general, the ex-situ material shows a mildly oblate distribution, flattened along the Z-direction 
\citep[e.g.][Monachesi et al. 2017]{2012MNRAS.420.2245M,2016MNRAS.459L..46M}. Only galaxies 
Au-2, Au-8, and Au-24 show a significantly flattened distribution that visually resembles the structure expected for a stellar disc. 
In some cases, such as Au-14, Au-19 and Au-20, clear signatures of cold substructure can be observed. This is debris from recent or on-going disruption events, which crosses the inner galactic regions, but has not yet had 
time to fully mix.

\begin{figure*}
\begin{flushleft}
    \includegraphics[width=170mm,clip]{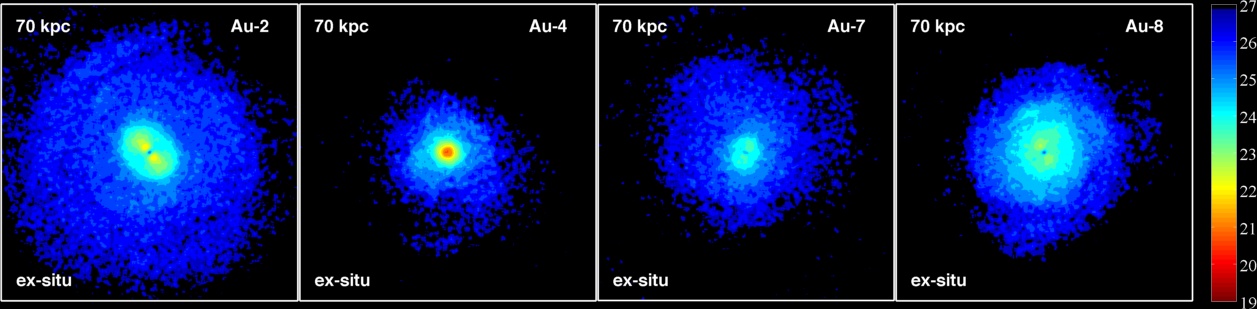}\\
    \includegraphics[width=170mm,clip]{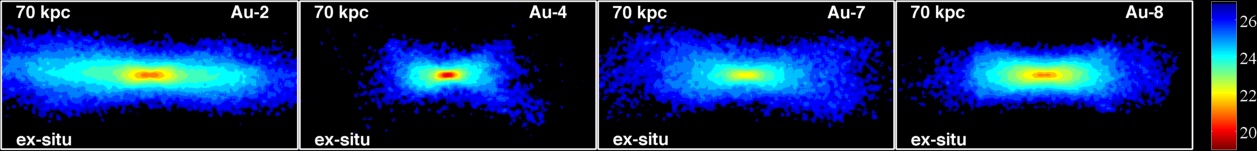}\\
    \includegraphics[width=170mm,clip]{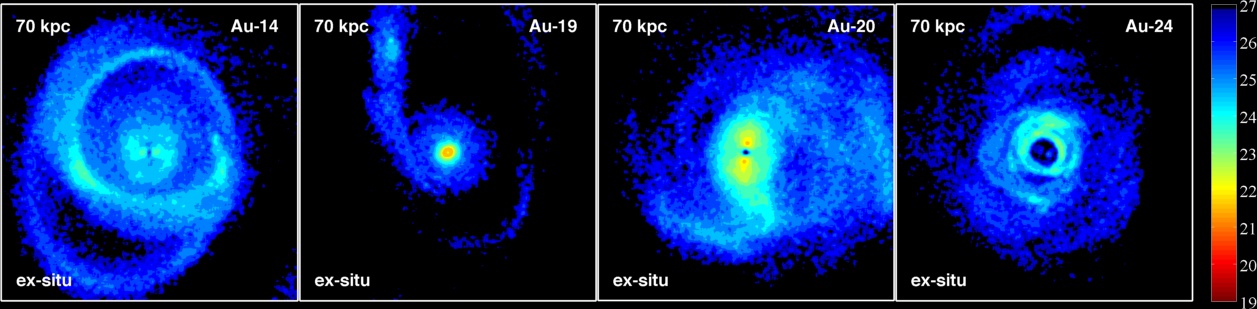}\\
    \includegraphics[width=170mm,clip]{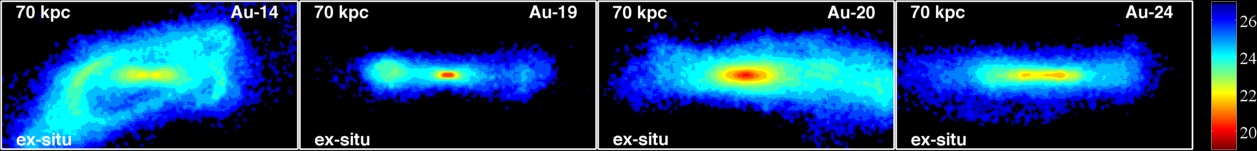}
    \end{flushleft}
    \caption{As in Figure~\ref{fig:sb_all} but for ex-situ star particles with circularity parameter $\epsilon \geq 0.7$. Note that,
    even though the side length of each panel is in all cases 70 kpc, only star particles located within  $R < R_{\rm opt}$ and $|Z| < 10$ kpc are considered as members of the ex-situ discs.}
    \label{fig:sb_eps07}
\end{figure*}

Figure~\ref{fig:sb_eps07} shows B-band surface brightness maps of the same galaxies, now obtained using only ex-situ star particles 
that satisfy the condition $\epsilon \geq 0.7$. This figure exposes, in all cases, a clear ex-situ disc component
composed of star particles on near-circular orbits. As discussed before, some of these ex-situ discs (e.g. Au-14 and Au-19),
show several spatially coherent stellar streams, associated with recent accretion events. On the other hand, discs such as Au-2 
show a very smooth spatial distribution. As we show below, the Au-2 ex-situ disc formed as a result of two $\sim 1:10$ mergers
that took place 8 Gyr ago, giving enough time for the resulting debris to fully mix.  

\begin{figure*}
    \includegraphics[width=160mm,clip]{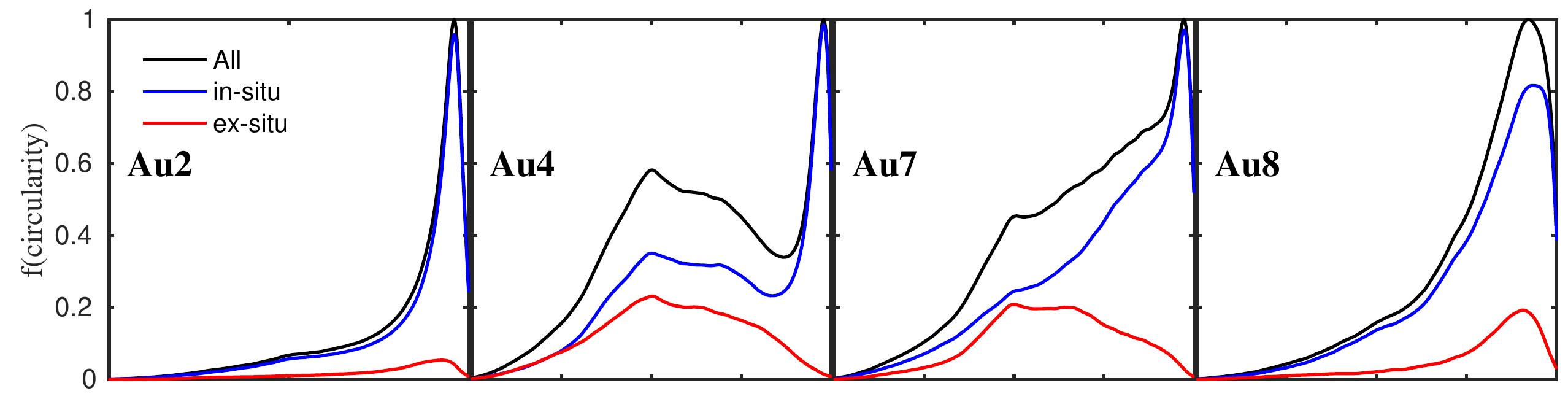}\\
    \includegraphics[width=160mm,clip]{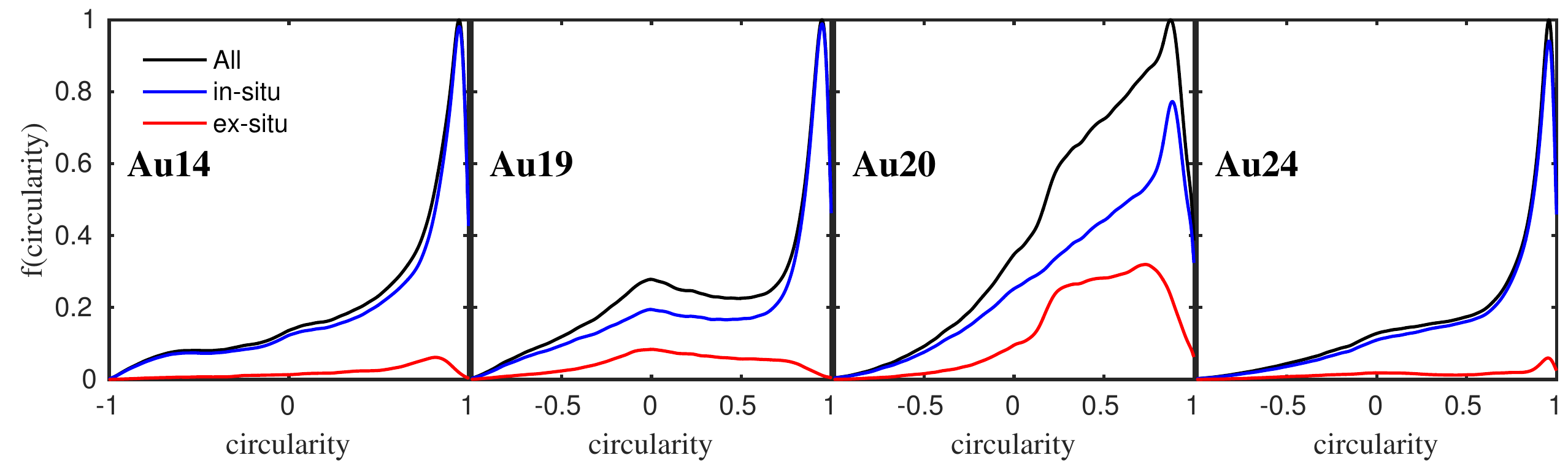}
    \caption{The black lines show the distribution of circularity parameter, $\epsilon$, for all star particles that are located within 
    $R < R_{\rm opt}$ and $|Z| < 10$ kpc in galaxies with a significant ex-situ disc. The blue and red lines show the contribution from the in-situ and ex-situ stellar populations, 
    respectively.}
    \label{fig:circus}
\end{figure*}

In Figure~\ref{fig:circus} we show the circularity ($\epsilon$) distribution of all star particles (black lines) that are located within our 
spatial disc selection box, i.e. $R<R_{\rm opt}$ and $|Z| < 10$ kpc, obtained from galaxies with significant ex-situ disc components. 
The blue and red lines show the contribution from the in-situ and ex-situ stellar populations, respectively. Interestingly, in half of our sample, the circularity distribution of the ex-situ component peaks at values of $\epsilon < 0.7$ 
(Au-4, Au-7, Au-19 and Au-20). However, for the remaining half, it peaks at $\epsilon \geq  0.7$. Note that, as discussed
in Section~\ref{sec:intro},  previous studies that attempted to identify an ex-situ component in the Milky Way disc have focused 
their analysis on stellar samples with $0.2 < \epsilon < 0.8$ (R14, R15). This selection
criterion is clearly justified by the complexity
of detecting an ex-situ component on in-situ dominated near-circular orbits ($\epsilon > 0.7$). However, this figure shows 
that most stars of a hypothetical accreted Milky Way disc could be buried in this region.

\section{Formation of ex-situ discs}
\label{sec:formation}

In this section we explore how and when these ex-situ discs are formed. In particular, we are interested in characterizing 
the number, orbital properties, and the total and stellar mass spectrum of the satellites
that have contributed to their formation.  

\begin{figure}
    \includegraphics[width=85mm,clip]{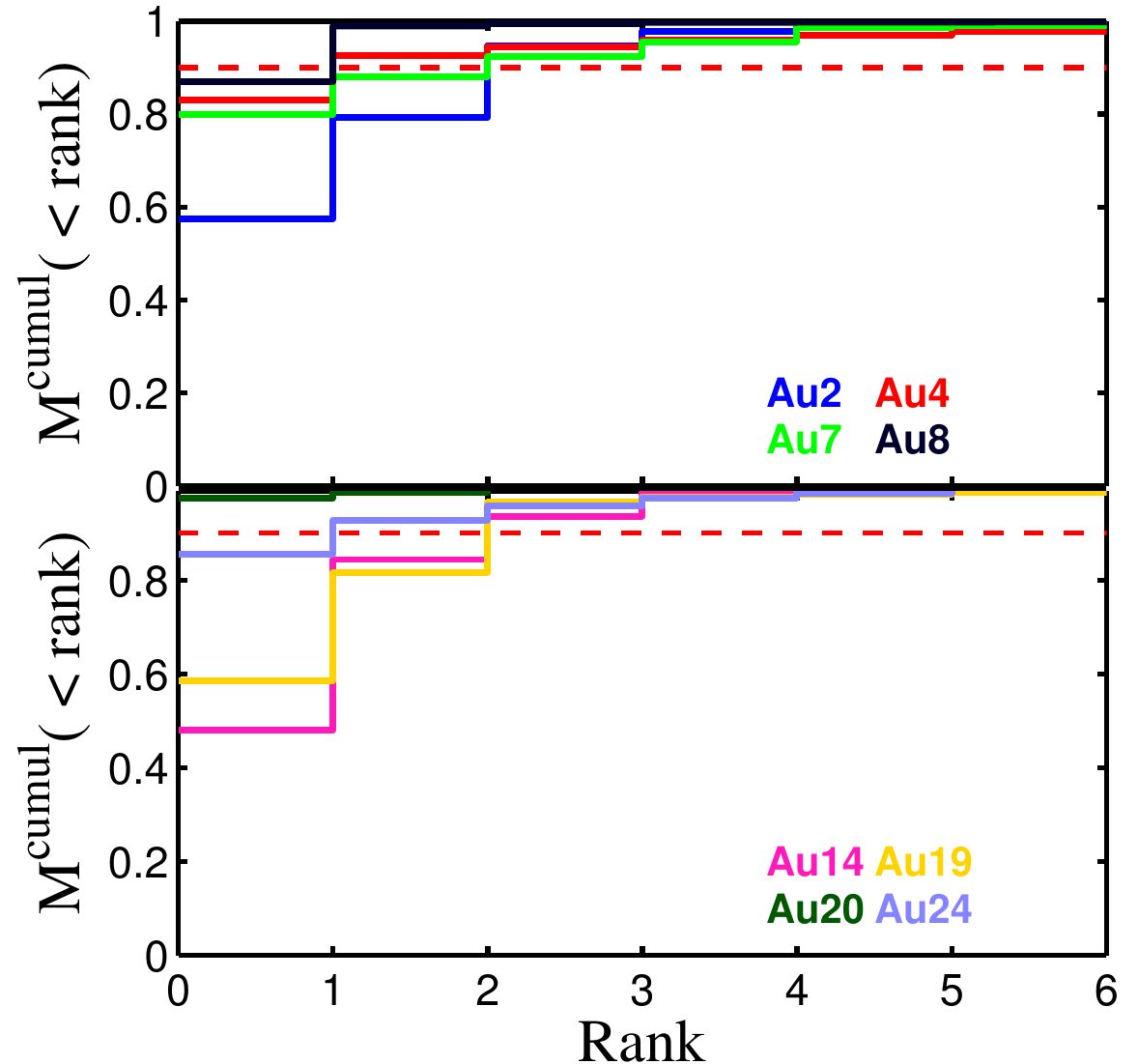}
    \caption{Cumulative stellar mass fractions obtained from each significant ex-situ disc. The step-like cumulative fractions show the contribution from the six most massive contributors to each ex-situ disc in rank order of decreasing mass. The red dashed line shows the 90 per cent level.}
    \label{fig:sig_prog}
\end{figure}

In Figure~\ref{fig:sig_prog} we decompose the 
total ex-situ disc mass into different accreted satellite contributors. Satellites 
are ranked according to their fractional mass contribution in decreasing order (i.e. the larger the contributor, 
the smaller the rank assigned). Interestingly, we find that only a few satellites are needed to account for the
bulk of the mass of the ex-situ discs. The number of 
significant contributors, defined as the  number of satellites required to account for $90\%$ of the mass, 
ranges from 1 to 3, with a median of 2. 
This is smaller than  the number of significant contributors that build up the accreted stellar halos of these galaxies, 
which ranges between 3 to 8, with a median of 5 \citep[][Monachesi et al. 2017]{cooper}. Note as well that, in all 
cases, there is a dominant contributor that accounts for $\gtrsim 50\%$ of the total mass.   

In Table~\ref{t2} we list some of the main properties of the most significant contributor to each ex-situ disc. Second 
 contributors are  listed in those cases where their contribution is also significant ($\gtrsim 20\%$ of the ex-situ disc mass). In several cases, we find that the 
 most significant contributors are massive satellites. Their peak total masses, i.e.  the maximum instantaneous
mass that these satellites have reached, are $M_{\rm sat}^{\rm peak} > 10^{11}~M_{\odot}$. Thus, they are 
associated with large merger events. However, we also find cases in which these satellites 
 have relatively low mass. The peak masses of the significant contributors range from 
 $0.6 \times 10^{11}~M_{\odot}$ (Au-24)  to $5.3 \times 10^{11}~M_{\odot}$ (Au-4). As
 expected,  second significant contributors are associated with less massive satellites. We note that, in all 
 these cases,  $M_{\rm peak}$ is achieved just prior to infall through the host virial radius.

Table~\ref{t2} also lists the lookback time, $t_{\rm cross}$, at which each satellite first crosses the host virial radius, $R_{\rm vir}$,
for the first time. The dispersion in $t_{\rm cross}$ values is large, ranging from early infall events with 
$t_{\rm cross} = 9.1$ Gyr (Au-14) to very late infall events with $t_{\rm cross} = 3.1$ Gyr (Au-4). Au-4 is an interesting 
case in which the host undergoes a major merger of mass ratio $M_{\rm sat}/ M_{\rm host} \approx 0.67$,  
approximately 3 Gyr ago. The host disc is strongly perturbed but survives the interaction and quickly regrows (within 2 Gyr) 
to reach a present-day  optical radius of $R_{\rm opt} = 24.5$ kpc.

Finally, Table~\ref{t2} lists the satellite's infalling angle, $\theta_{\rm infall}$, defined as the angle between the 
disc angular momentum  and that of the satellite's orbit, both measured  at $t_{\rm cross}$. Again we 
find a large spread in $\theta_{\rm infall}$, with values that range from $15^{\circ}$ (Au-20) to $85^{\circ}$ (Au-7).
Significant ex-situ discs are expected to form from mergers events in which massive satellites are accreted with 
low grazing angles, such as Au-20. Ex-situ discs formed as a result of mergers with massive satellites 
that are accreted with large $\theta_{\rm infall}$ are more interesting and thus we study them in more detail.  

\begin{figure*}
    \includegraphics[width=49.5mm,clip]{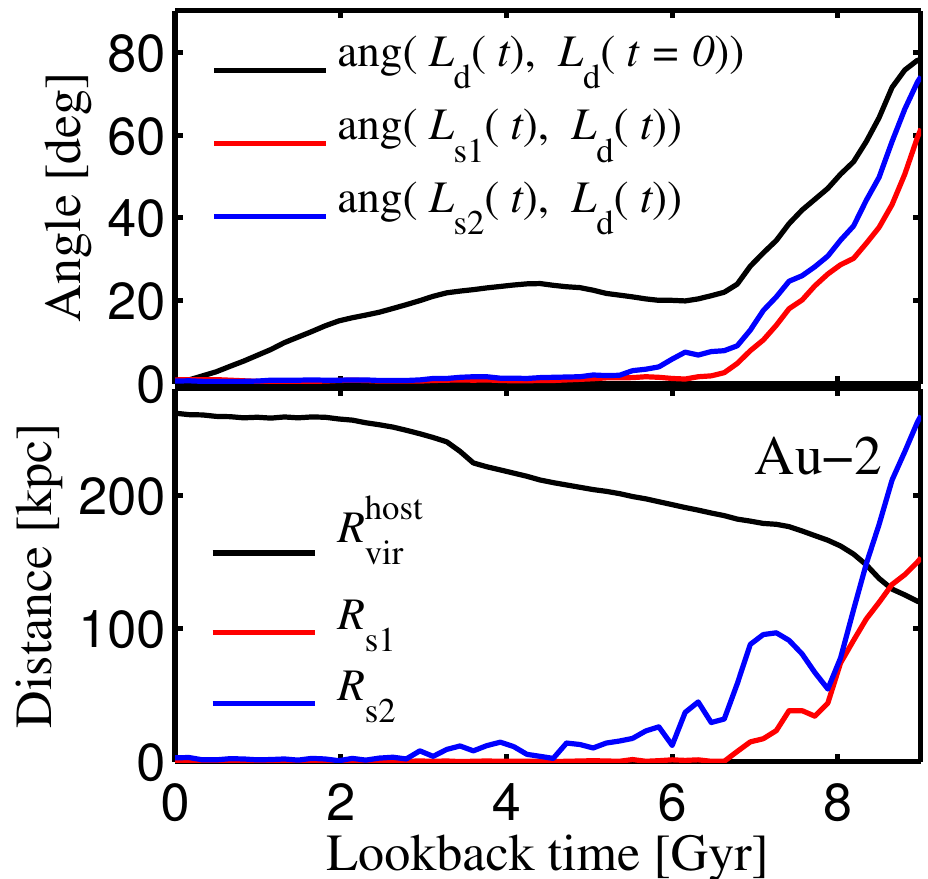}
    \includegraphics[width=41mm,clip]{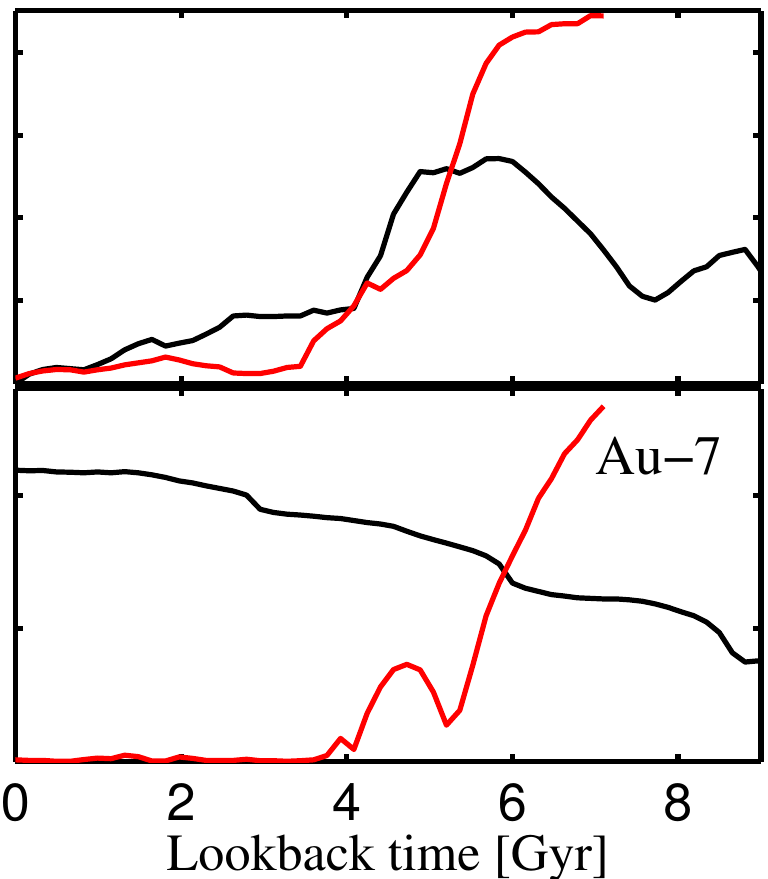}
    \includegraphics[width=41mm,clip]{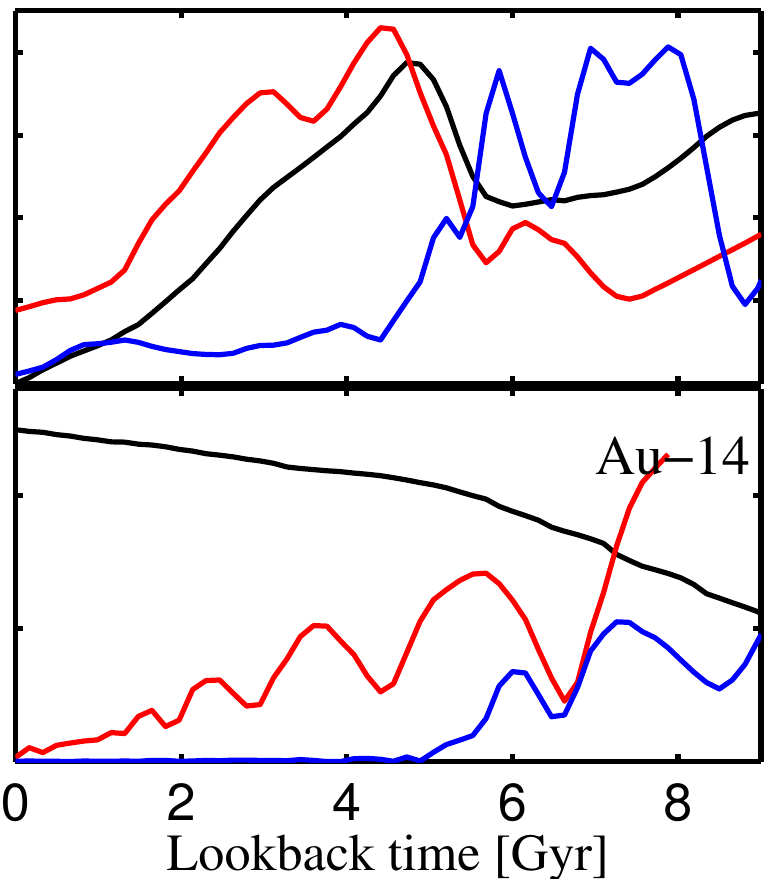}
    \includegraphics[width=41mm,clip]{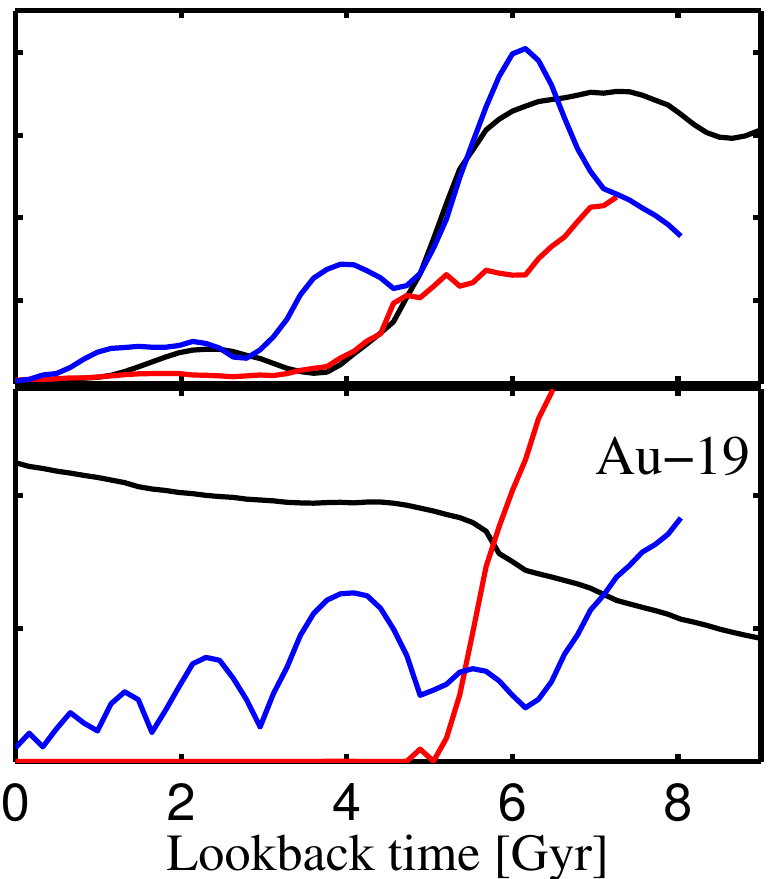}
    \caption{Top panels: The red (blue) lines show the time evolution of the angle between the angular momentum 
    of the disc and that of the orbit of the most significant (second) contributor. The black lines show the time 
    evolution of the orientation of the angular momentum vector of the disc with respect to its orientation at the present-day. 
    Bottom panels: The red (blue) lines show the time evolution of the  galactocentric distance of the 
    most significant (second) contributor. The black lines show the time evolution of the host $R_{\rm vir}$.}
    \label{fig:ang}
\end{figure*}

In the top panels of Figure~\ref{fig:ang} we show with red (blue) lines the time evolution of the angle between the 
disc angular momentum 
and the most significant (second) contributor orbital angular momentum vectors. Only four representative examples 
are shown, but similar results are found for the remaining galaxies. It is interesting to notice how 
these massive satellites start to align with the host disc as soon as, or even before, they 
cross the host $R_{\rm vir}$. This is particularly clear in Au-2. The angle between the disc and the two most 
significant contributors  before they cross $R_{\rm vir}$, $\sim 9$ Gyr ago,  is $\sim 60^{\circ}$ and $70^{\circ}$, 
respectively. For reference, we show in the bottom panels of  Fig.~\ref{fig:ang} the time evolution of 
the satellites' galactocentric distances, $R_{\rm s1}$ and $R_{\rm s2}$, and of the host 
virial radius, $R_{\rm vir}$. In many cases, it only takes $\sim 2$ Gyr for these satellites to become almost perfectly aligned 
with the host
disc. 

A very similar situation can be seen in the remaining examples. Note that the rapid alignment between these two angular momentum 
vectors is due not only to changes in the orbits of the satellite galaxies, but also to a strong response
of the host galactic discs. This can be seen in the top panels of Fig.~\ref{fig:ang} where we show, with black lines, the 
time evolution of the disc's angular momentum vector orientation with respect to its orientation at the present-day. In general, 
the discs start to rapidly tilt as soon as the satellites cross $R_{\rm vir}$, and this continues until the satellites are 
fully merged. For example, the Au-2 disc tilts $\sim 60^{\circ}$ during the merger of the two ex-situ disc most significant 
contributors. As discussed in \citet[][and references therein]{2016MNRAS.456.2779G}, 
even low-mass satellites that 
penetrate the outer regions of a galaxy can significantly perturb and tilt a host galactic disc. This is not only
due to direct tidal perturbation \citep[e.g.][]{2008MNRAS.391.1806V,2015MNRAS.452.2367Y} 
but also to the generation of asymmetric features in the DM halo  that can be efficiently 
transmitted to its inner regions, thereby affecting the deeply embedded  disc.

Note that the two most significant contributors  to the ex-situ disc in Au-2 
are accreted onto the host as a group, but they are not bound to each other. 
This can be seen from their very similar $t_{\rm cross}$ and $\theta_{\rm infall}$ values (Table~{\ref{t2}}). 
In the remaining galaxies with two significant contributors, the satellites are independently accreted.

\begin{table}
\centering
\caption{Main properties of the most significant contributors to the ex-situ discs. The columns are 1) Model name, 2) Satellite's 
peak total mass, 3) Lookback time at which each satellite crosses the host $R_{\rm vir}$ for the first time, and 4) Angle between the 
angular momentum vectors of the disc and the satellite orbit, measured  at $t_{\rm cross}$.}
\label{t2}
\begin{tabular}{c c c c}
\hline
Run & $M_{\rm sat}^{\rm peak}$ & $t_{\rm cross}$ & $\theta_{\rm infall}$ \\
 & $[10^{11}~M_{\odot}]$ & [Gyr] & \\
\hline
\hline
Au-2 & 1.2 & 8.6 & $45^{\circ}$ \\
 & 0.9 & 8.3 & $45^{\circ}$ \\
\hline
Au-4 & 5.3 & 3.1 & $60^{\circ}$ \\
\hline
Au-7 & 1.8 & 6 & $85^{\circ}$ \\
\hline
Au-8 & 1.0 & 8 & $25^{\circ}$ \\
\hline
Au-14 & 1.0 & 7.3 & $45^{\circ}$ \\
 & 0.8 & 9.1 & $30^{\circ}$ \\
\hline
Au-19 & 2.9 & 5.8 & $25^{\circ}$ \\
 & 0.6 & 7.1 & $47^{\circ}$ \\
\hline
Au-20 & 2.5 & 6 & $15^{\circ}$ \\
\hline
Au-24 & 0.6 & 8.6 & $80^{\circ}$ \\
\hline
%Au-30 & 1.7 & 7.3 & $72^{\circ}$ \\
\end{tabular}
\end{table}

\section{Characterization of  ex-situ discs}
\label{sec:charac}

\begin{figure*}
    \includegraphics[width=58mm,clip]{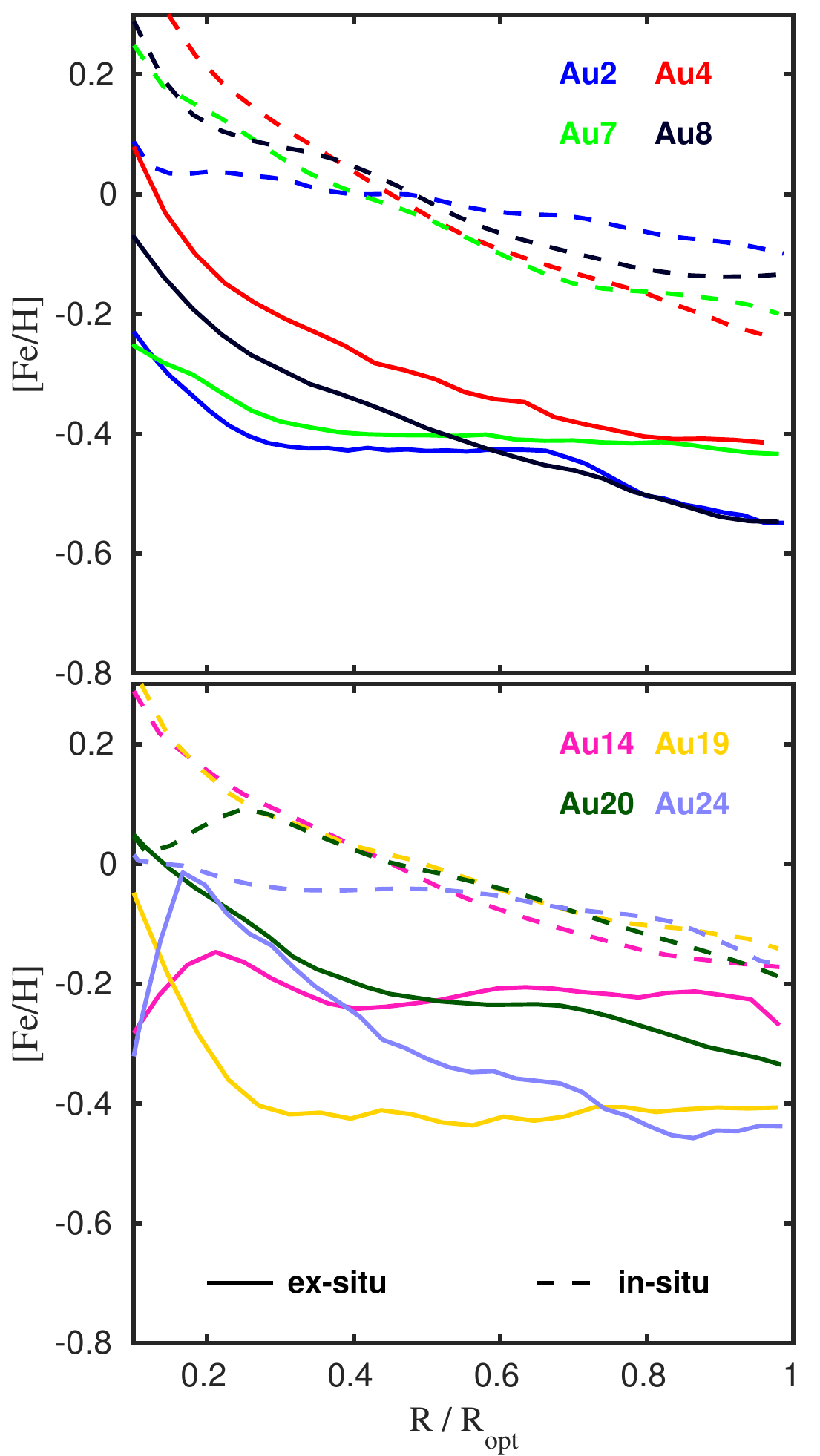}
    \includegraphics[width=58mm,clip]{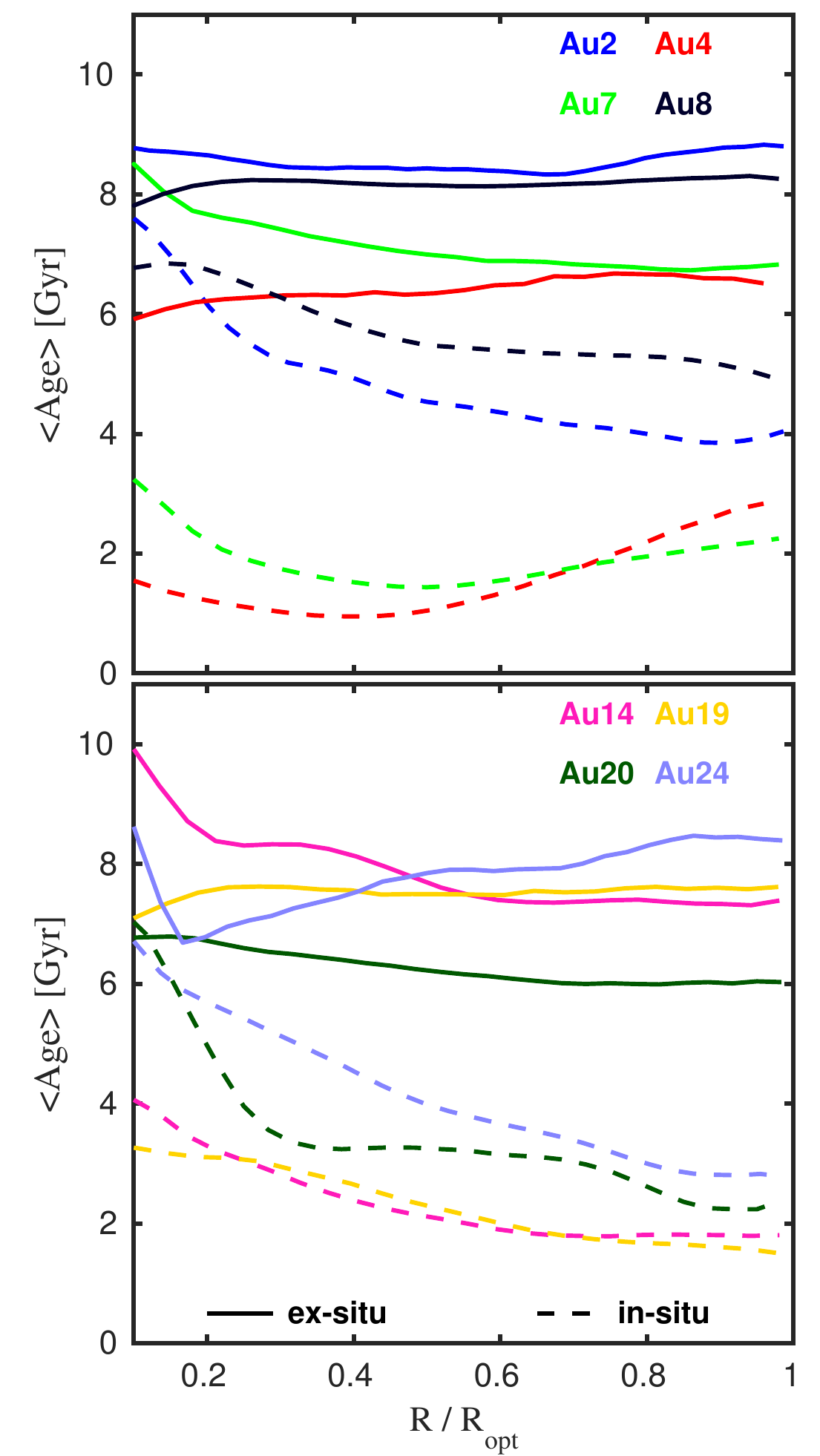}
    \includegraphics[width=58mm,clip]{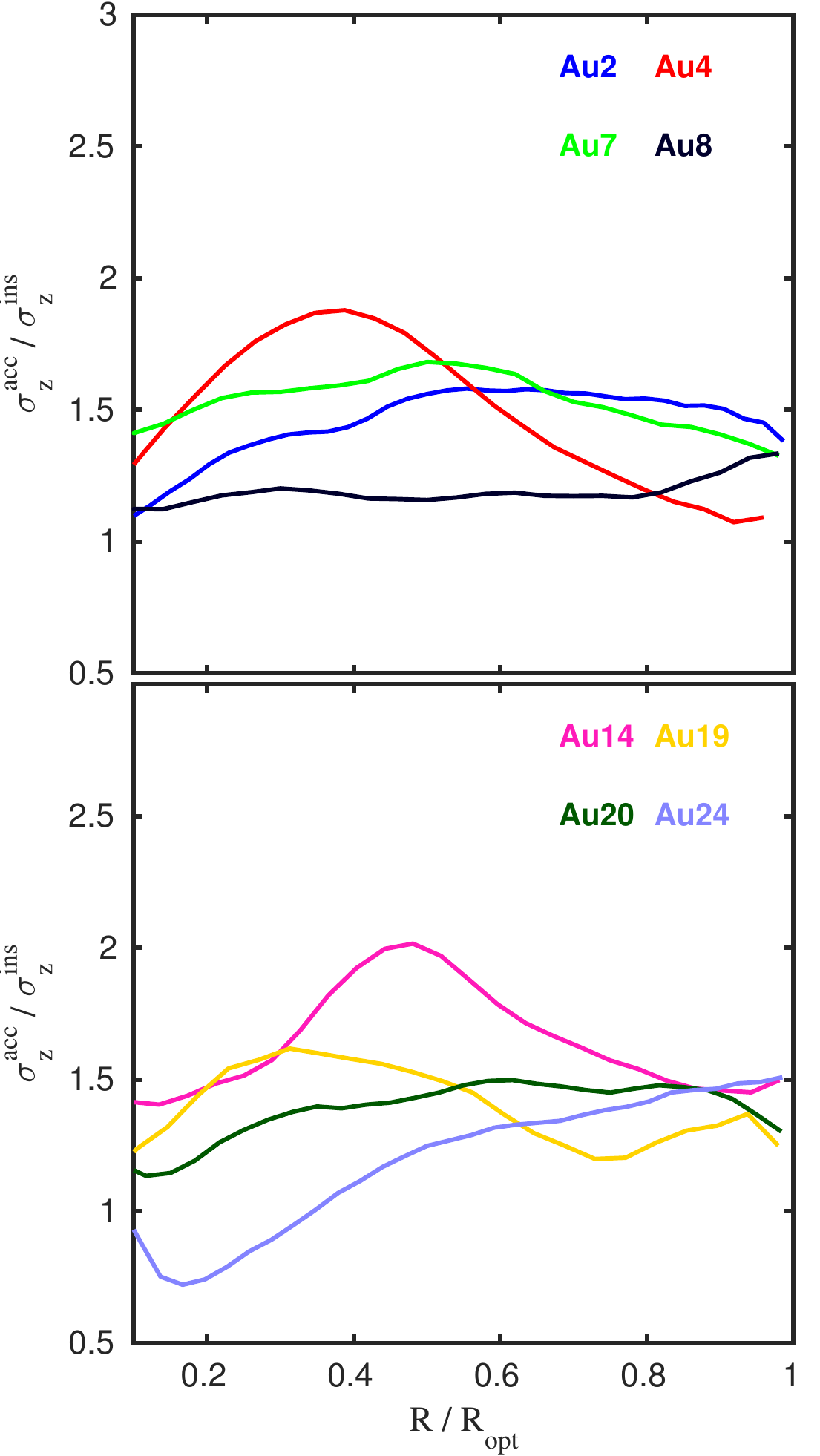}
    \caption{Left and middle panels: Median [Fe/H] and age profiles for the discs as a function of galactocentric distance. 
    The solid lines show the profiles obtained from the ex-situ stellar populations whereas the dashed lines give those obtained from their in-situ 
    counterparts. Right panels: ratio of the ex-situ to in-situ mass-weighted vertical dispersions as a function of galactocentric distance.}
    \label{fig:props}
\end{figure*}

In this Section we characterize the main stellar population properties and the vertical distribution of each 
ex-situ disc. The left panels of Figure~\ref{fig:props} show median [Fe/H] profiles as a function of galactocentric 
distance. The solid lines are the profiles obtained from the ex-situ stellar populations whereas the
dashed lines are those obtained from their in-situ counterparts. In all cases, the in-situ [Fe/H] profiles show clear negative 
[Fe/H] radial gradients, associated with inside-out formation of the main disc
\citep[e.g.][GR17]{2001ApJ...554.1044C,2006MNRAS.366..899N, 2014A&A...572A..92M}. In general, ex-situ discs
 also exhibit [Fe/H] gradients. These are a reflection of the [Fe/H] gradients of the most significant contributors prior to their disruption. Note
 that the galaxy with the most metal-rich and steepest [Fe/H] gradient in the ex-situ disc component is Au-4. As previously discussed, the 
 largest contributor to this ex-situ disc is a $\sim 5.3 \times 10^{11} M_{\odot}$ satellite that crossed the host $R_{\rm vir}$ 
 just $\sim 3$ Gyr ago.
 
 A detailed analysis of the chemical evolution of the Auriga galaxies will be presented in a forthcoming work. Here we are mainly interested
in the differences between the in-situ and ex-situ [Fe/H] profiles. We find that, in all cases and at all radii, 
ex-situ discs are significantly more metal-poor than their in-situ counterparts. 
Differences in median metallicity can be as large as $\sim 0.5$ dex (Au-2, Au-19). 

The middle panels of Figure~\ref{fig:props} show the median age profiles as a function of galactocentric 
distance for both stellar 
populations. In all cases,  ex-situ discs (solid lines) show approximately flat age profiles, reflecting the 
median age of the stellar populations of the most significant satellite contributor. Note that ex-situ discs with younger 
populations accreted their most significant contributor later on (Au-4, Au-7 and Au-20).
Median ages of ex-situ 
disc populations range between 6 and 9 Gyr.  Conversely, the mean age profiles of the in-situ stellar 
discs show, in general, negative
gradients, reflecting their inside-out formation. In addition, this component is significantly younger than the ex-situ disc, with differences in 
median ages as large as $\sim 6$ Gyr (e.g. Au-14).

Finally, the right panels of Figure~\ref{fig:props} show  the ratio of  ex-situ to in-situ  
mass-weighted vertical dispersion, $\sigma_{\rm Z}$ as a function of galactocentric distance. This quantity provides a measure of
disc thickness. In general, we find ex-situ discs to be thicker than the 
in-situ components, with typical values of  
$\sigma_{\rm Z}^{\rm exsitu} / \sigma_{\rm Z}^{\rm insitu} \sim 1.5$. 
Some galaxies show a clear negative gradient  in the outer disc regions
($R \gtrsim 0.5 R_{\rm opt}$, e.g. Au-4, Au-7 and Au-14). 
As shown by GR17 and \citet{2017MNRAS.465.3446G}, the in-situ component of these Auriga discs shows strong
flaring, warping and bending in the outer regions. This causes $\sigma_{\rm Z}^{\rm insitu}$ to rise steeply at 
galactocentric distances where the disc stops being strongly cohesive due to its weak self-gravity.

The significant differences that the ex-situ and in-situ disc stellar populations exhibit could be used to define 
indicators to identify ex-situ discs. We discuss this further in Section~\ref{sec:identy}.

\section{Identification of ex-situ discs}

\label{sec:identy}

In the previous Section we have shown that the median age and [Fe/H] of the ex-situ disc stellar populations are 
older and more metal-poor, respectively, than their in-situ counterparts. Here we explore how these two 
characteristics could be used to search for this galactic component in our own
Galactic disc. 

\begin{figure*}
    \includegraphics[width=85mm,clip]{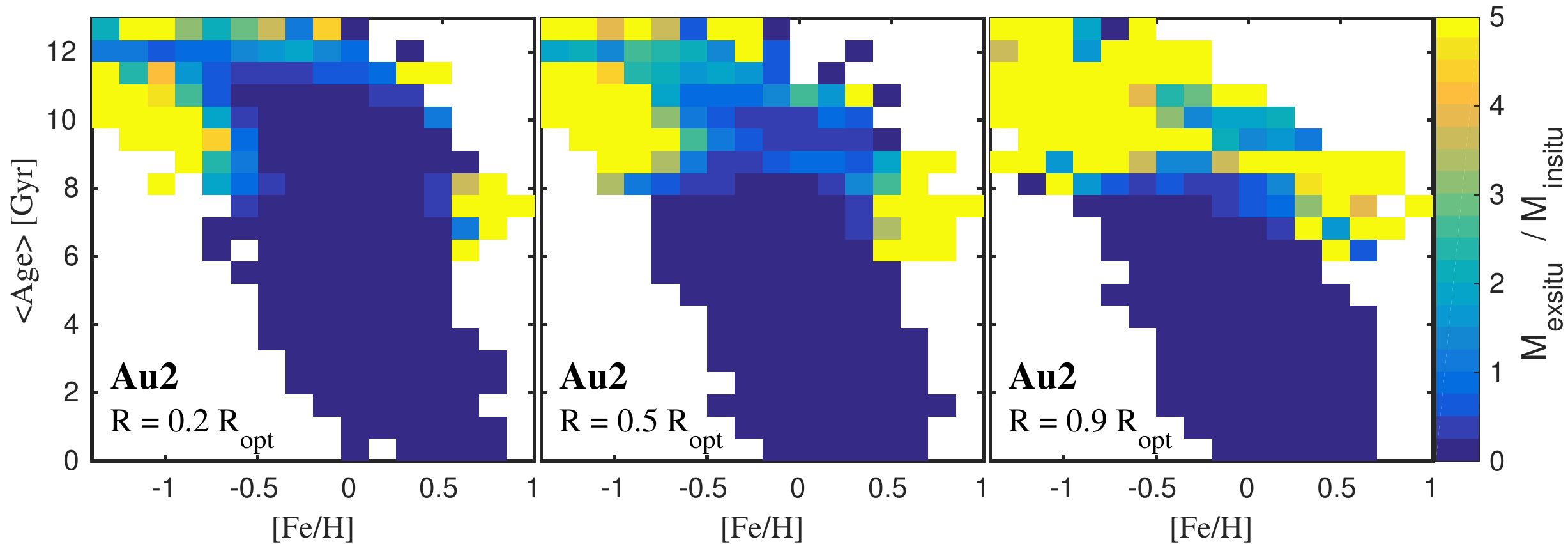}
    \includegraphics[width=85mm,clip]{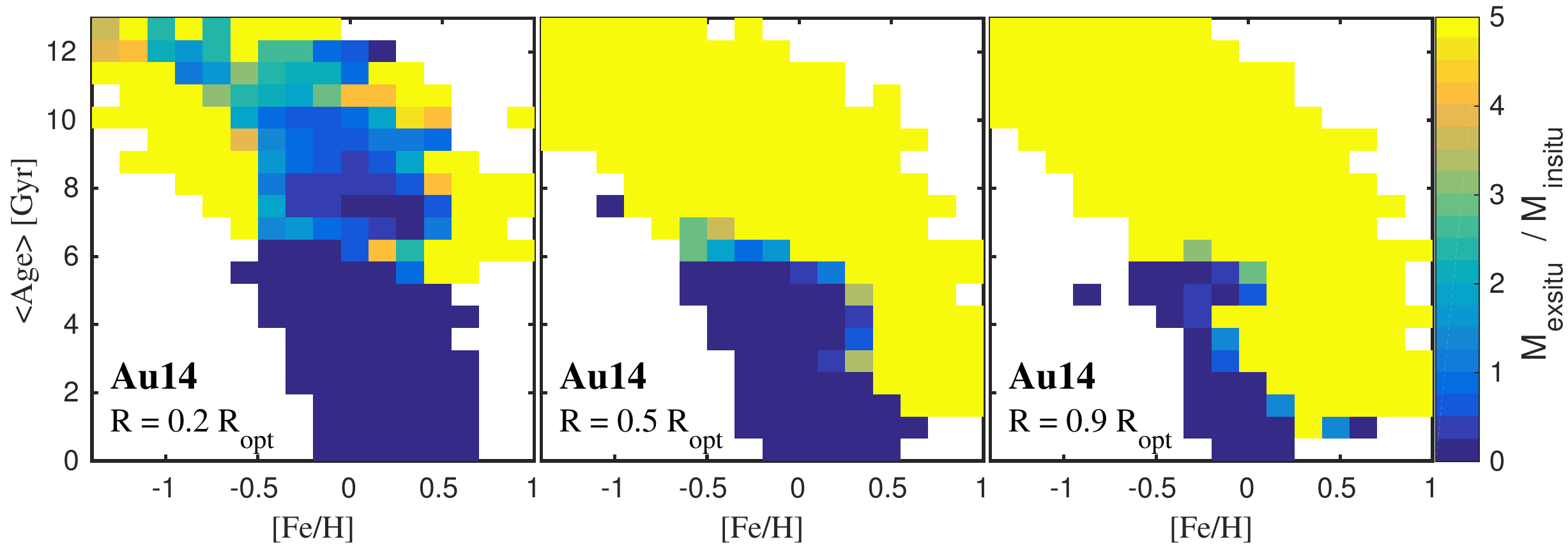}\\
    \includegraphics[width=85mm,clip]{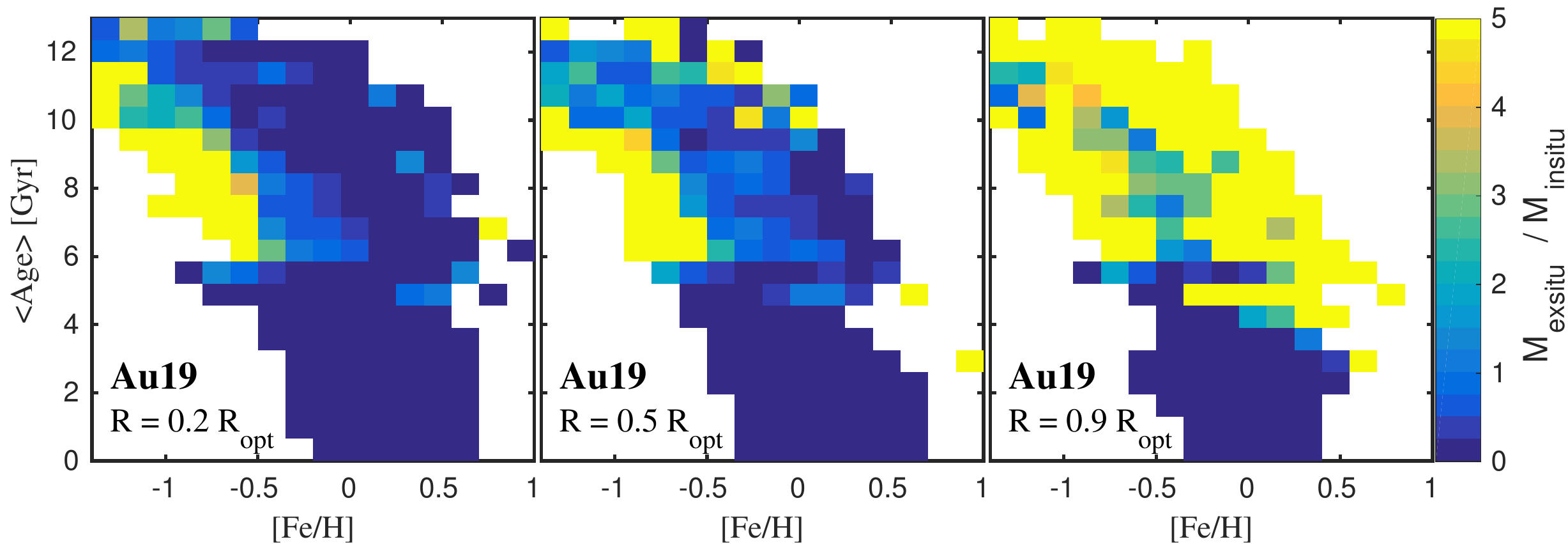}
    \includegraphics[width=85mm,clip]{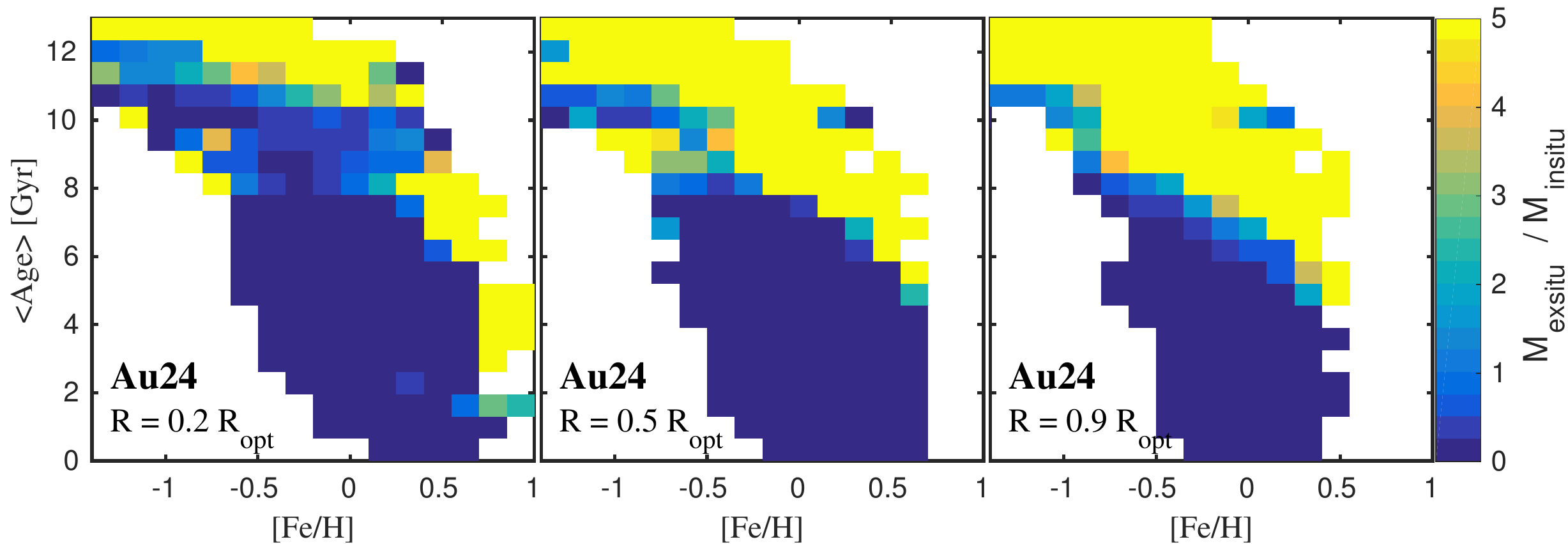}
    \caption{Two dimensional histograms of the ratio of ex-situ to in-situ disc stellar mass, $\nu = M_{\rm exsitu}/M_{\rm insitu}$,
    in age and [Fe/H] space, obtained from four representative
    galaxies with significant ex-situ discs. For each galaxy three panels are shown. Each panel shows the results obtained from the 
    stellar particles located within three different cylindrical shells defined as $(0.2 \pm 0.1,~0.5 \pm 0.1,~0.9 \pm 0.1)R_{\rm opt}$. 
    Regions dominated by in-situ ($\nu < 1$) and ex-situ ($\nu \geq 5$) disc components are shown in dark blue and yellow,
    respectively.}
    \label{fig:age-feh}
\end{figure*}

In Figure~\ref{fig:age-feh} we show the ex-situ to in-situ mass ratio, $\nu$, obtained from subsets of stellar 
particles located in different regions of the age and [Fe/H] space. As before, we only show four representative 
examples, but similar results are found for the remaining galaxies with significant ex-situ discs.
To generate these two dimensional histograms
we first selected all disc stellar particles (recall, $\epsilon \geq 0.7$, $R < R_{\rm opt}~{\rm and}~|Z| < 10$ kpc) 
located within three different
cylindrical shells defined as $(0.2 \pm 0.1,~0.5 \pm 0.1,~0.9 \pm 0.1)R_{\rm opt}$. On each cylinder we 
gridded the (Age, [Fe/H]) space with an $N \times N$ regular mesh of bin size $(0.65~{\rm Gyr},~0.15~\rm{dex})$. 
Finally, we computed the ratio $\nu$ considering only the stellar particles that are located within each 
(Age, [Fe/H]) bin. 

The colour bar in Fig.~\ref{fig:age-feh} indicates  different values of $\nu$. Regions of the (Age-[Fe/H]) space that are dominated by in-situ stellar populations, i.e. $0 \leq \nu < 1$, are shown in dark blue.  
It is evident that, at all galactocentric distances, the in-situ disc  dominates in regions with young
and metal-rich stellar populations. Close to the galactic center, $0.2R_{\rm opt}$, ex-situ stellar 
populations  are  found to dominate at very old ages ($\gtrsim 8$ Gyr) and low [Fe/H] 
($\lesssim -0.5$ dex). Nonetheless, an interesting pattern arises when larger galactocentric distances 
are considered. It is clear from this figure that the regions dominated by the ex-situ disc
gradually increase as we move further out. In a few examples (Au-14 and Au-24), 
 regions with ages $> 6$ Gyr are mainly dominated by ex-situ stellar populations already at $0.5R_{\rm opt}$. 
Note that, assuming an optical 
radius of 19 kpc for the Milky Way \citep{2017arXiv170107831L}, these regions can be regarded as Solar 
Neighborhood analogs. The trend continues for larger galactocentric distances ($0.9R_{\rm opt}$). This indicates
that the likelihood of identifying an ex-situ disc in samples of old stars increases towards the outskirts of the
galactic discs. The reason for this was already discussed in Section~\ref{sec:charac}. In general, the discs in
this subset of Auriga galaxies show inside-out formation (see Fig.~\ref{fig:props}). The in-situ
stellar populations become, on average, younger with galactocentric distance. Conversely, we find that the age
distribution of the ex-situ population remains nearly constant with galactocentric distance. Thus, as the 
in-situ populations recede towards regions with younger ages, the ex-situ population takes over.   

\begin{figure*}
    \includegraphics[width=85mm,clip]{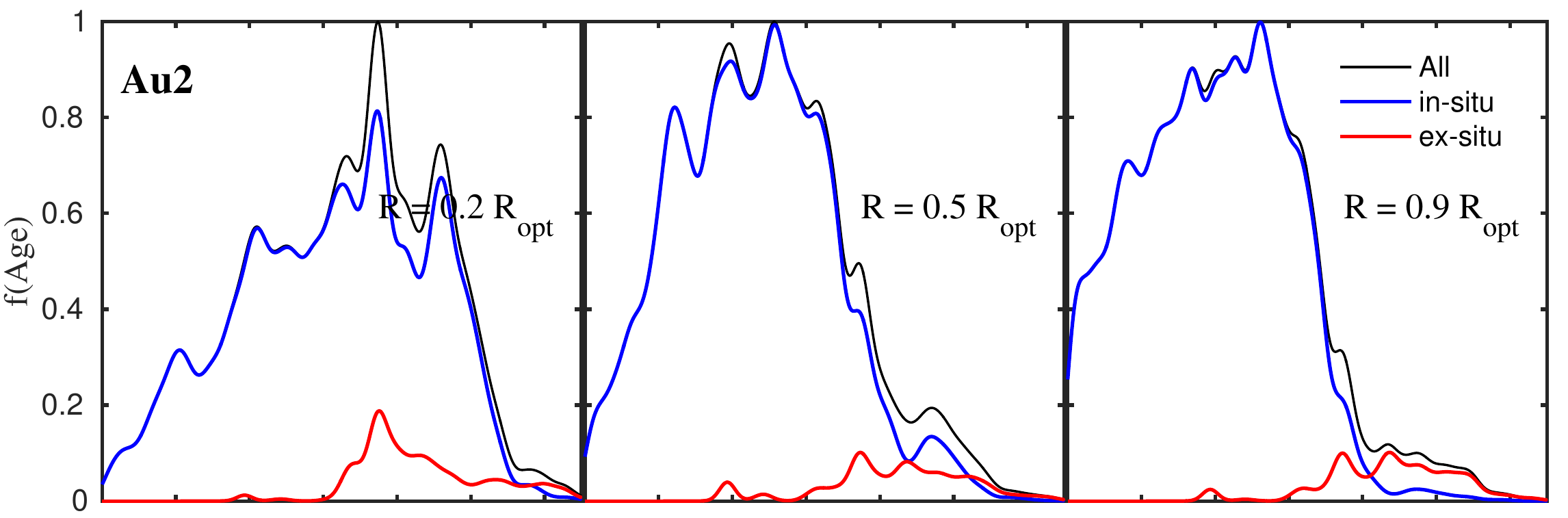}
    \includegraphics[width=85mm,clip]{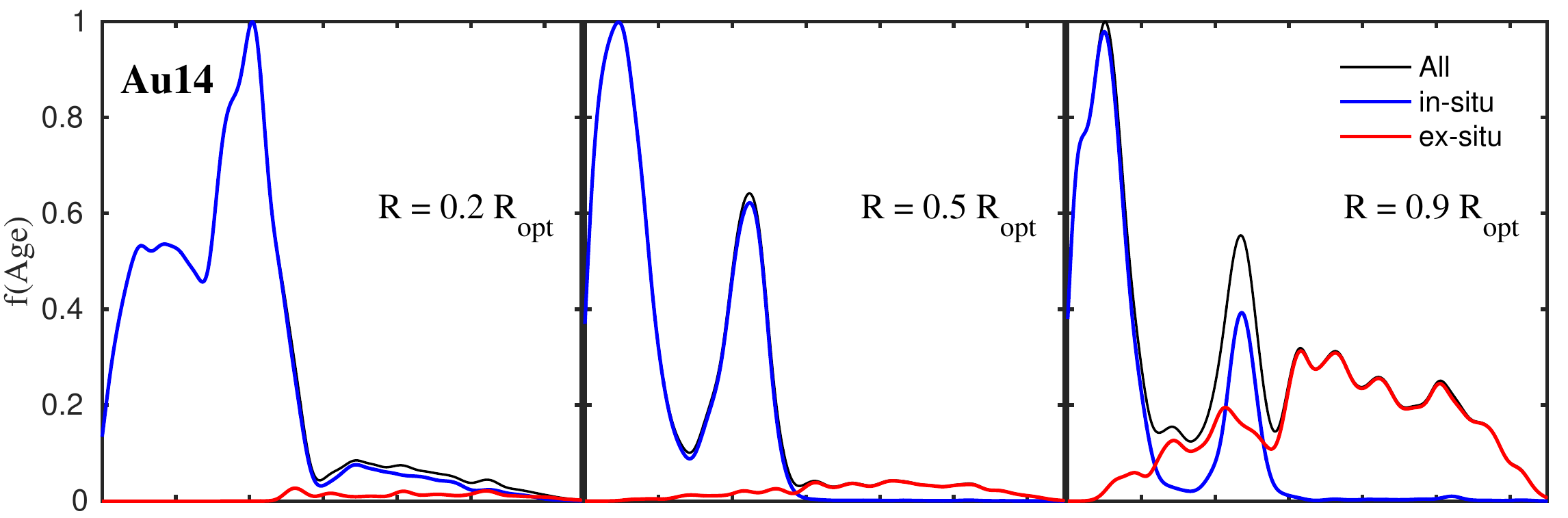}\\
    \includegraphics[width=85mm,clip]{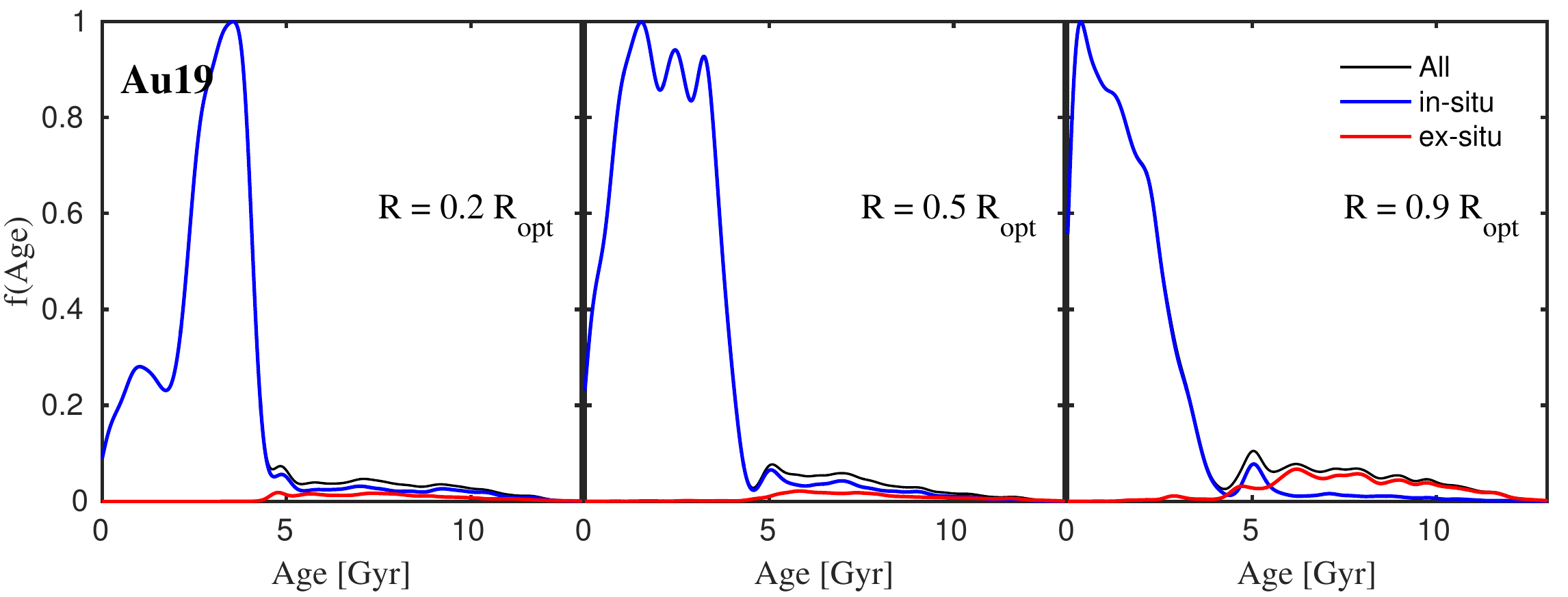}
    \includegraphics[width=85mm,clip]{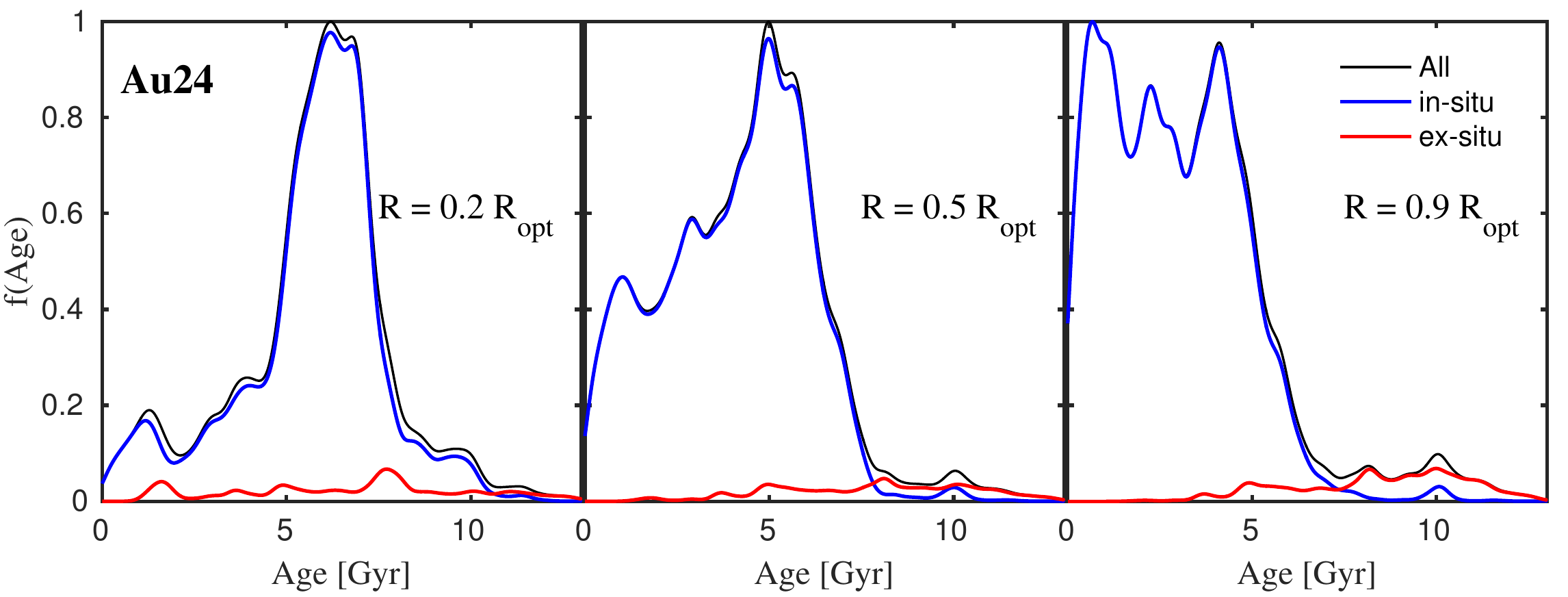}
    \caption{Stellar age distribution of disc particles obtained within the three cylindrical shells used in Figure~\ref{fig:age-feh}.
    The red and blue lines show the distributions obtained from the in-situ and ex-situ stellar particles, respectively. The black lines show the overall
    distributions.}
    \label{fig:age}
\end{figure*}

This can be seen more clearly in Figure~\ref{fig:age}, where we show Gaussian kernel histograms of the stellar
age distribution for the four disc examples previously discussed. Note that no cut in [Fe/H] has been 
imposed to generate these histograms. Again, we can see that close to the galactic center, 
i.e. $\sim 0.2 R_{\rm opt}$, the in-situ stellar populations (blue lines) dominate these distributions at all ages.
As we move outwards, the ex-situ populations (red lines) start to take over at old ages. In all cases, the old
tail of these distributions (age $\gtrsim 7$ Gyr) is dominated by ex-situ star particles at galactocentric distances of $\sim 0.9R_{\rm opt}$. 

It is important to highlight that the presence of an old tail in this distribution does not necessarily 
imply the presence of an ex-situ disc component. For example, in-situ old stars can be found in the outer regions of a
galactic disc as a result of processes such as radial migration \citep[][and references therein]{2016MNRAS.459..199G}. 
To unambiguously identify ex-situ 
stars additional information based on chemical tagging should be used (R14, R15).

\begin{figure}
    \includegraphics[width=85mm,clip]{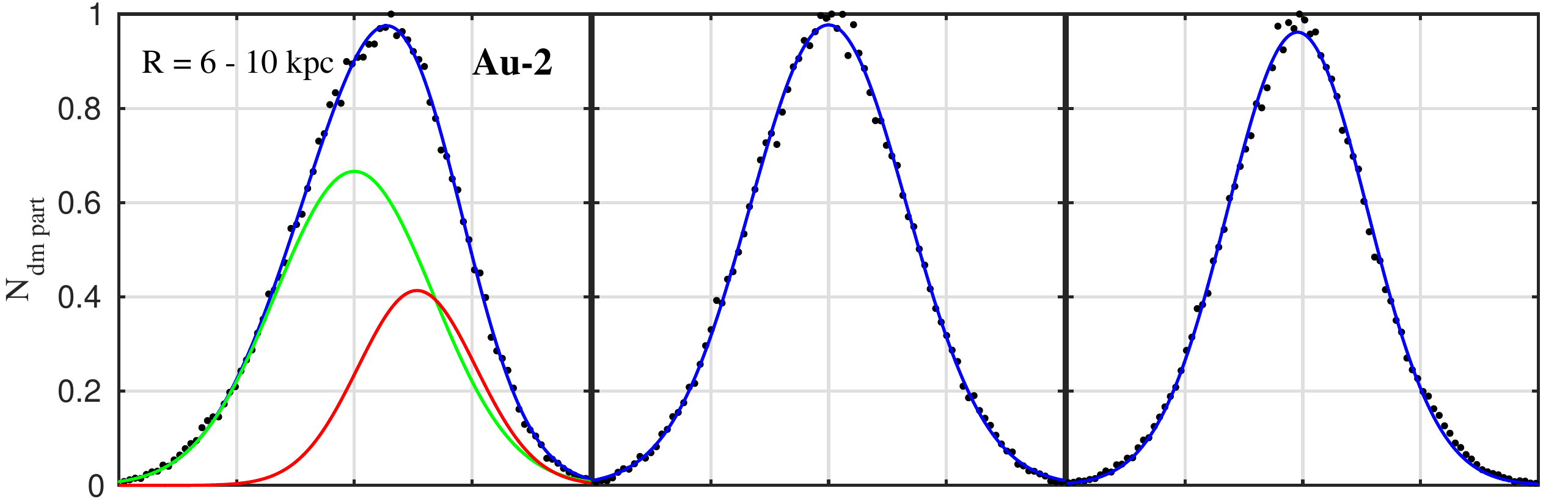}\\
    \includegraphics[width=85mm,clip]{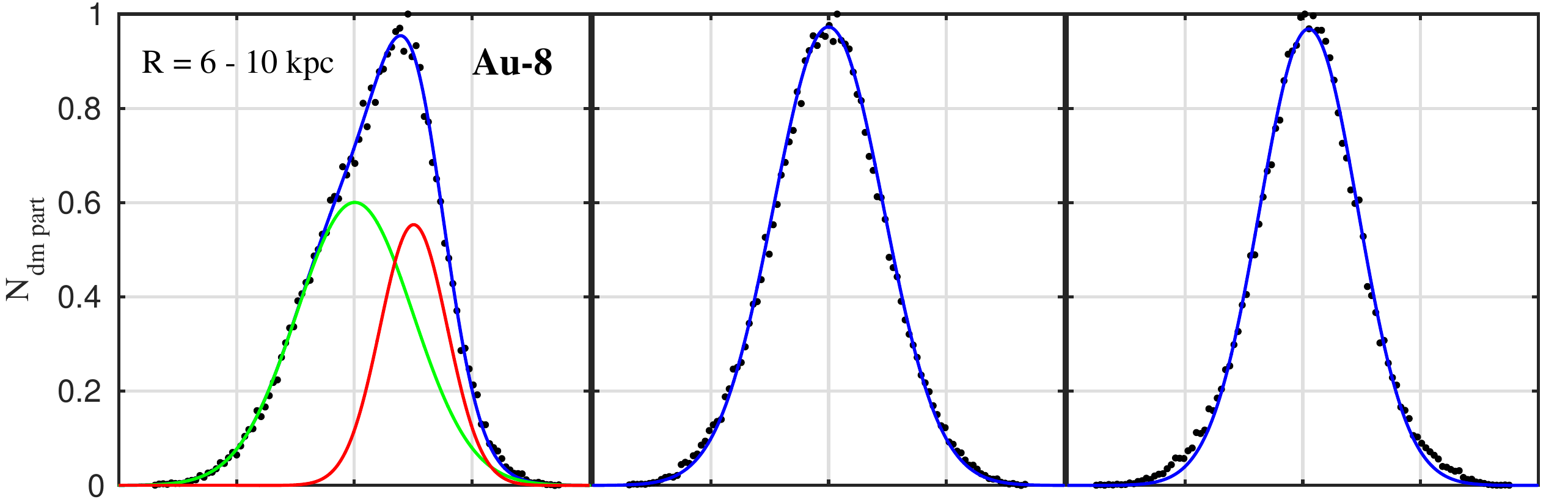}\\
    \includegraphics[width=85mm,clip]{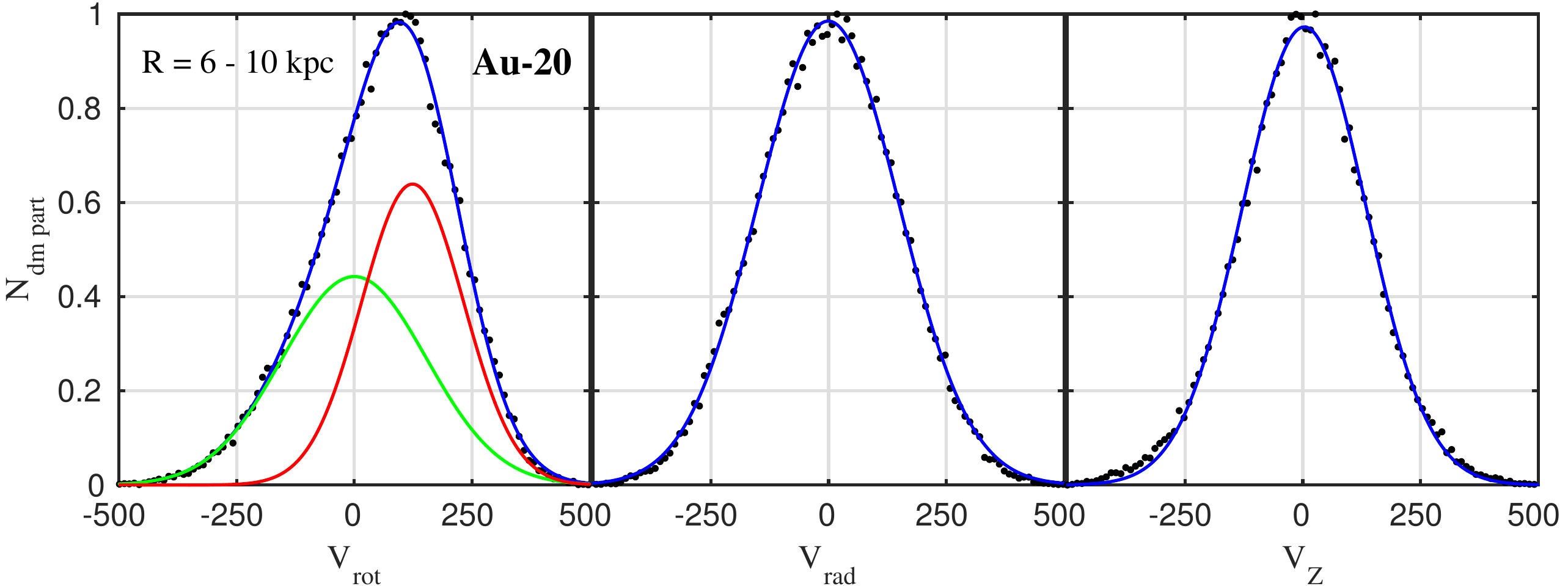}
    \caption{The black dots indicate the velocity distributions of the dark matter particles located within a cylinder defined as 
    $|Z| \leq 5$ kpc and $6 \leq R \leq 10$ kpc. The left, middle and right panels show the rotational, radial and vertical velocity 
    distributions, respectively. The blue lines in the left panels show the result of a double Gaussian fit to the data. 
    The red and green lines show the contributions from each individual Gaussian. The blue lines in the other two panels show the result of single Gaussian fits to the corresponding data.}
    \label{fig:dm_disc}
\end{figure}

\section{Discussion}
\label{sec:discussion}

In this work we have shown that massive satellites can be can be  disrupted in a plane that is well-aligned with that of the 
host disc, depositing material on near-circular orbits  that are dynamically indistinguishable from those 
of stars born in-situ. An ex-situ disc would not only be relevant for probing the merger history of
the Milky Way, but would also hint at the presence of an underlying DM disc 
\citep[see e.g.][]{2008MNRAS.389.1041R,2009ApJ...703.2275P,2009MNRAS.397...44R}. 
The quantification and characterization of the DM discs in the Auriga simulations will be presented in another paper 
(Schaller et al., in prep). Here we merely show that 
some of the Auriga galaxies with significant ex-situ discs also have a significant rapidly rotating DM component. 

In Figure~\ref{fig:dm_disc} we show three examples of the velocity distribution of dark matter particles located within 
a cylinder defined as $|Z| \leq 5$ kpc and $6 \leq R \leq 10$ kpc. In all cases we can see that both the radial, $V_{\rm rad}$,
and the vertical, $V_{\rm z}$, velocity components can be well fitted with a single Gaussian  centred at 0 km s$^{-1}$. 
However, it is clear that a single Gaussian centred at 0 km s$^{-1}$ cannot describe the azimuthal velocity distributions, $V_{\rm rot}$. 
Following \citet{2009MNRAS.397...44R}, we used a double Gaussian to describe these
distributions. As in \citet{2016MNRAS.461L..56S}  one of the Gaussian is centred at $V_{\rm rot} = 0$ km s$^{-1}$. 
Very similar results are obtained if the center of both Gaussian are left as free parameters. The blue lines show the result of such 
fits, while the red and green lines show the contribution from each individual Gaussian. In all cases we find  the  second Gaussian
to be centred at high $V_{\rm rot}$, with values of $\sim 133$, 124 and 126 km s$^{-1}$ for Au-2, Au-8 and Au-20, respectively. Following \citet{2016MNRAS.461L..56S}, we estimate the amount of DM that the secondary Gaussian contributes to these cylinders by evaluating
its integral. We find contributed mass fractions of $32\%,~35\%$ and $50\%$, respectively. In general, galaxies with
significant ex-situ discs show $V_{\rm rot}$ distributions that cannot be described with single Gaussian (Schaller et al. in prep).

The Au-2 case is particularly interesting. It possesses a significant rotating DM component and, as previously 
discussed, the circularity distribution of its ex-situ component (within the spatially defined disc) peaks 
at  $\epsilon \geq 0.7$ (see also Au-8). The  most significant contributors were accreted $\sim 8.5$ Gyr ago, and  
quickly merged with the host. More importantly, this galaxy has a smoothly rising age-vertical velocity 
dispersion relation during the last 8 Gyr of evolution \citep[see][figure 15]{2016MNRAS.459..199G}, and thus 
its behaviour is qualitatively consistent with that observed in the MW \citep{2011A&A...530A.138C,2014ApJ...781L..20M}. 
Recently, \citet[][D17]{2017arXiv170309230D} combined Gaia data release 1 astrometry with Sloan Digital Sky Survey (SDSS) images 
to measure proper motions of old stars in the MW stellar halo. They find a gently rotating prograde signal, which shows little 
variation with Galactocentric radius out to 50 kpc. As discussed by D17, some Auriga galaxies with significant ex-situ discs (Au-2,
Au-4, Au-7, Au-19 and Au-24) also show mildly rotating old stellar halos, consistent with the observations.  
An ex-situ component with the characteristics found in e.g. Au-2, and its associated rotating DM component, would not have been detected to date in the MW disc.

\section{Conclusions}
\label{sec:conclusions}

We have studied the formation of ex-situ discs in model galaxies with similar mass to the 
MW, simulated in a fully cosmological context. An important goal of this study was to explore whether a significant 
ex-situ stellar component could be buried within the near-circular orbit population of the MW disc which is strongly 
dominated by in-situ stars. For this purpose, we  focused our analysis 
on star particles with large circularity parameter,  $\epsilon \geq 0.7$. This differs from the strategy in observational studies, 
such as those presented by R14 and R15. These attempted to identify an ex-situ disc in the Milky Way by focusing on stars with 
$0.2 < \epsilon < 0.8$.

Our study shows that approximately one third of our sample (8 out of 26 models) 
contains a significant ex-situ disc. These galaxies show an ex-situ
to in-situ disc ($\epsilon \geq 0.7$) mass ratio $\eta > 0.05$. Note however, that the fraction of models with significant 
ex-situ discs would be larger if we 
were to relax our circularity threshold to lower values. In fact, as shown in Figure 1 of R15, the circularity distributions of the 
ex-situ stellar discs presented in \citet[][]{2008MNRAS.389.1041R} peak at $\epsilon \sim 0.5$. 
We find that, in general, the ex-situ to 
in-situ disc mass fraction rises as we move towards the outer disc regions. 

We have characterized the circularity distribution of all stellar particles that are spatially located within regions 
associated with the galactic discs (i.e. $R < R_{\rm opt}$ and $|Z| < 10$ kpc). Half of the ex-situ discs sample have a distribution that is 
consistent with those shown in R15, peaking at values of $\epsilon < 0.7$. Interestingly, for the remaining half we find a
circularity distribution that peaks at values of $\epsilon \geq 0.7$. Such discs would not have been detected in existing
observational studies.

In general, ex-situ discs are formed from the debris of fewer than three massive satellite galaxies, but most of their mass ($>50 \%$)
always comes from a single significant contributor. The peak total mass of this dominant contributor ranges between  $6 \times 10^{10}~M_{\odot}$ 
and $5.3 \times 10^{11}~M_{\odot}$. Both the virial radius crossing time and the infall angle of these satellites have a very large scatter,
with values ranging between 3.1 to 9.1 Gyr and $15^{\circ}$ to $85^{\circ}$, respectively. We highlight
that significant ex-situ discs can arise from merger events with massive satellites that are accreted at infall 
angles as large as $85^{\circ}$ \citep[see also][]{2009MNRAS.397...44R}. In these cases we find that the disc and satellite angular momentum 
vectors rapidly align. This is not  purely due to an evolution of the infalling satellite's 
orbit, but also to a strong response of the host galactic disc. We find that host discs start to tilt as soon as these 
massive satellites cross $R_{\rm vir}$. This tilt can be driven both by
direct tidal perturbations \citep[e.g.][]{2008MNRAS.391.1806V,2015MNRAS.452.2367Y,2017MNRAS.465.3446G}
and by the generation of asymmetric features in 
the host DM halo that can extend into the inner regions, affecting the deeply embedded  disc 
\citep[][and references therein]{2016MNRAS.456.2779G}.
It is important to note that the response of a disc depends, among other things, on its vertical 
rigidity. A disc tilts as a whole  only in regions where it is 
strongly cohesive thanks to its self-gravity \citep[e.g.][]{2006MNRAS.370....2S}. Thus, the frequency and properties 
of ex-situ discs may be misrepresented in simulations of Milky 
Way-like galaxies if these are not able to reproduce correctly the vertical structure of 
the MW disc \citep[see e.g.][]{2010MNRAS.406..576P}. This could be an issue for the Auriga simulations, since our final  
stellar discs are thicker than observed ($h_{\rm z} \sim 1$ kpc, see GR17). 
Hence, their vertical rigidity may be significantly lower than that of the MW disc.

Ex-situ discs tend to be  thicker than in-situ discs, with vertical dispersion ratios  
$\sigma_{\rm Z}^{\rm exsitu} / \sigma_{\rm Z}^{\rm insitu} \sim 1.5$. In all cases, and at all radii, the 
ex-situ disc component is significantly more metal-poor than the in-situ disc. Differences
in the median [Fe/H] can be as large as 0.5 dex. Ex-situ discs are also significantly older than their in-situ counterparts, with 
age differences that can be as large as 6 Gyr. Their median age profiles are flat and reflect the median
age of the stellar populations from the most significant contributors. In contrast, the median age profiles of the in-situ discs show, 
in general, negative gradients, reflecting the inside-out formation of these stellar discs. 

We have shown that the different properties that the in-situ and ex-situ stellar populations exhibit could be 
used to isolate ex-situ star candidates on near-circular orbits in the Milky Way disc (recall, $\epsilon \geq 0.7$). 
By gridding the Age-[Fe/H] space, and computing
the ex-situ to in-situ mass ratio within each bin, we find that the regions dominated by ex-situ disc gradually increase as we move 
towards the outer disc, while in the inner galactic regions (i.e. 
$R \sim 0.2R_{\rm opt}$)  in-situ stellar populations dominate nearly everywhere. In a few cases we find that regions with ages $> 6$
Gyr are dominated by ex-situ disc stars already at galactocentric distances $R = 0.5R_{\rm opt}$. 
 The  likelihood of identifying
an  ex-situ  disc  in  samples  of old  stars  increases  towards  the
outskirts of the disc. However, the presence of an old tail in the age
distribution may  not uniquely imply  the presence of an  ex-situ disc
component. To unambiguously identify  such an ex-situ component in the
MW,  additional  information  based  on  chemical  tagging  should  be
considered.

\section*{Acknowledgements}
We are grateful to Adrian Jenkins and David Campbell for the selection
of the sample and making the initial conditions. RG and VS acknowledge
support through the DFG Research Centre SFB-881 `The Milky Way System'
through project A1. VS and RP acknowledge support by the European
Research Council under ERC-StG grant EXAGAL-308037. This work was
supported by the Science and Technology Facilities Council (grant
number ST/L00075X/1) and the European Research Council (grant number
GA 267291, `Cosmiway'). This work used the DiRAC Data Centric system
at Durham University, operated by the Institute for Computational
Cosmology on behalf of the STFC DiRAC HPC Facility (www.dirac.ac.uk).
This equipment was funded by BIS National E-infrastructure capital
grant ST/K00042X/1, STFC capital grants ST/H008519/1 and ST/K00087X/1,
STFC DiRAC Operations grant ST/K003267/1 and Durham University. DiRAC
is part of the National E-Infrastructure. SB acknowledges support from
the International Max-Planck Research School for Astronomy and Cosmic
Physics of Heidelberg (IMPRS-HD) and financial support from the
Deutscher Akademischer Austauschdienst (DAAD) through the program
Research Grants - Doctoral Programmes in Germany (57129429).

\bibliographystyle{mnras}
\bibliography{exsitu_disc}

\label{lastpage}

\end{document}